%% file: BPH-13-003_temp.tex
\pdfoutput=1

\documentclass[11pt,twoside,a4paper,cmspaper,final,collab]{cms-tdr}
\def\svnVersion{218305}\def\cmsCernNo{2013-140}\def\cmsCernDate{\today}\def\cmsMessage{Published in Physics Letters B as \href{http://dx.doi.org/10.1016/j.physletb.2013.10.055}{\doi{10.1016/j.physletb.2013.10.055.}}}

\begin{document}\cmsNoteHeader{BPH-13-003}

\hyphenation{had-ron-i-za-tion}
\hyphenation{cal-or-i-me-ter}
\hyphenation{de-vices}
\cmsNoteHeader{BPH-13-003}

\newcommand{\costh}{\ensuremath{\cos\vartheta}}
\newcommand{\lth}{\ensuremath{\lambda_\vartheta}}
\newcommand{\lph}{\ensuremath{\lambda_\varphi}}
\newcommand{\ltp}{\ensuremath{\lambda_{\vartheta\varphi}}}
\newcommand{\ltilde}{\ensuremath{\tilde{\lambda}}}
\newlength{\digitwidth} \settowidth{\digitwidth}{0}
\newcommand{\range}[2]{\ensuremath{^{#2}_{#1}}}
\newcommand{\fbg}{\ensuremath{f_{\PB}}}
\newcommand{\fNP}{\ensuremath{f_\mathrm{NP}}}
\newcommand{\fbgtot}{\ensuremath{f_{\PB,\text{tot}}}}
\def\Tiny{\fontsize{5pt}{5pt}\selectfont}
\newcommand{\PsiOne}{\JPsi} % \JPsi has built-in xspace
\newcommand{\PsiTwo}{\Pgy\xspace}
\newcommand{\PsiN}{\ensuremath{\psi\cmsSymbolFace{(nS)}}\xspace}
\newcommand{\mlab}[1]%
{\mbox{}\marginpar[\raggedleft\hspace{0pt}#1]{\raggedright\hspace{0pt}#1}}

\RCS$Revision: 213052 $
\RCS$HeadURL: svn+ssh://svn.cern.ch/reps/tdr2/papers/BPH-13-003/trunk/BPH-13-003.tex $
\RCS$Id: BPH-13-003.tex 213052 2013-10-22 13:56:38Z carlosl $
\newlength\cmsFigWidth
\ifthenelse{\boolean{cms@external}}{\setlength\cmsFigWidth{0.45\textwidth}}{\setlength\cmsFigWidth{0.64\textwidth}}
\ifthenelse{\boolean{cms@external}}{\providecommand{\cmsLeft}{top}}{\providecommand{\cmsLeft}{left}}
\ifthenelse{\boolean{cms@external}}{\providecommand{\cmsRight}{bottom}}{\providecommand{\cmsRight}{right}}
\cmsNoteHeader{BPH-13-003}
\title{\texorpdfstring{Measurement of the prompt \PsiOne and \PsiTwo polarizations in pp collisions at $\sqrt{s} = 7$\TeV}{Measurement of the prompt J/psi and psi(2S) polarizations in pp collisions at sqrt(s) = 7 TeV}}

\date{\today}

\abstract{The polarizations of prompt \PsiOne and \PsiTwo mesons are
measured in proton-proton collisions at $\sqrt{s} = 7$\TeV, using a dimuon data sample
collected by the CMS experiment at the LHC,
corresponding to an integrated luminosity of 4.9\fbinv.
The prompt \PsiOne and \PsiTwo polarization parameters $\lambda_\vartheta$,
$\lambda_\varphi$, and $\lambda_{\vartheta\varphi}$, as well as the frame-invariant
quantity $\tilde{\lambda}$, are measured from the dimuon decay angular distributions
in three different polarization frames.
The \PsiOne results are obtained in the transverse momentum range
$14<\pt<70$\GeV, in the rapidity intervals $\abs{y} < 0.6$ and $0.6 < \abs{y} < 1.2$.
The corresponding \PsiTwo results cover $14<\pt<50$\GeV and include
a third rapidity bin, $1.2 < \abs{y} < 1.5$.
No evidence of large
polarizations is seen in these
kinematic regions, which extend much beyond those previously explored.}

\hypersetup{%
pdfauthor={CMS Collaboration},%
pdftitle={Measurement of the prompt J/psi and psi(2S) polarizations in pp collisions at sqrt(s) = 7 TeV},%
pdfsubject={CMS},%
pdfkeywords={CMS, physics, quarkonium production, quarkonium polarization}}

\maketitle

\section{Introduction}

After considerable experimental and theoretical efforts over the past decades, the
understanding of quarkonium production in hadron collisions is still
not fully settled~\cite{bib:QWG}.
In particular, the polarization of \PsiOne mesons is not
satisfactorily described in the context of nonrelativistic
quantum chromodynamics
(NRQCD)~\cite{bib:NRQCD}, where the purely perturbative
colour-singlet production~\cite{bib:lansberg-HP08} is complemented
by processes including possible nonperturbative transitions from
coloured quark pairs to the observable bound states. The \mbox{S-wave}
quarkonia directly produced at high transverse momentum, \pt, are
predicted to be transversely polarized~\cite{bib:BK,bib:Lei,bib:BKL}
with respect to the direction of their own momentum.
Contrary to this expectation, the CDF Collaboration measured a small
longitudinal polarization in prompt \PsiOne production~\cite{bib:CDFpolRun2}.
Since the measurement includes both directly produced \PsiOne mesons and
those resulting from feed-down decays of heavier charmonia, the comparison
between the theoretical predictions and the experimental results remained
ambiguous~\cite{bib:Faccioli-EPJC}.
Also the apparent lack of kinematic continuity between the fixed-target and the
collider quarkonium polarization data~\cite{bib:Faccioli-PRL-FT2Coll} raises
doubts on the reliability of these complex measurements.
Given the absence of feed-down decays from heavier charmonia affecting \PsiTwo
production, the measurements of the \PsiTwo polarization should be particularly informative,
especially if made with higher accuracy and extending up to higher \pt than those provided
by CDF~\cite{bib:CDFpolRun2}.

The polarization of the $J^\mathrm{PC} = 1^{--}$ quarkonium
states can be measured through the study of the
angular distribution of the leptons produced in their $\Pgmp\Pgmm$
decay~\cite{bib:Faccioli-EPJC},
\begin{linenomath}
\begin{equation}
\label{eq:angular_distribution}
\ifthenelse{\boolean{cms@external}}{
\begin{split}
& W(\costh,\varphi|\vec{\lambda}) \,
= \, \frac{3}{4\pi(3 + \lambda_{\vartheta})}
(1  +  \lambda_{\vartheta} \cos^2 \vartheta\, + \\
&
+ \lambda_{\varphi} \sin^2 \vartheta \cos 2 \varphi
+ \lambda_{\vartheta \varphi} \sin 2 \vartheta \cos \varphi ) \,,
\end{split}
}{%
W(\costh,\varphi|\vec{\lambda}) \,
= \, \frac{3}{4\pi(3 + \lambda_{\vartheta})}
(1  +  \lambda_{\vartheta} \cos^2 \vartheta\,
+ \lambda_{\varphi} \sin^2 \vartheta \cos 2 \varphi
+ \lambda_{\vartheta \varphi} \sin 2 \vartheta \cos \varphi ) \,,
}
\end{equation}
\end{linenomath}
where $\vartheta$ and $\varphi$ are the polar and azimuthal angles, respectively, of the
$\Pgmp$ with respect to the $z$ axis of the chosen polarization frame.
Robust quarkonium polarization measurements
require extracting all the angular distribution parameters,
$\vec{\lambda} = (\lth, \lph, \ltp)$, in at least two polarization
frames, as well as a frame-invariant polarization parameter,
$\tilde{\lambda} = (\lambda_\vartheta + 3 \, \lambda_\varphi) /
(1-\lambda_\varphi)$~\cite{bib:Faccioli-PRL-FrameInv,bib:Faccioli-PRD-FrameInv,bib:Faccioli-shapes}.
This approach was followed in the $\Upsilon$ polarization
analysis of CDF~\cite{PhysRevLett.108.151802}, in recent
theoretical calculations~\cite{Baranov:2011ib}, in the
detailed study of the \PgUa, \PgUb, and
\PgUc\ polarizations performed by
CMS~\cite{bib:UpsPol-CMS}, and in the recent
measurements of the \PsiOne\ polarization at forward
rapidity reported by ALICE~\cite{bib:psiPol-ALICE}
and LHCb~\cite{bib:psiPol-LHCb}.
This Letter presents the analogous measurement of the polarizations of the
\PsiOne and \PsiTwo mesons (abbreviated as \PsiN, with $n=1,2$)
promptly produced in pp collisions at a
centre-of-mass energy of 7\TeV, at the LHC. The analysis is based on a dimuon
sample collected in 2011, corresponding to an integrated luminosity of
4.9\fbinv.
The \PsiOne (\PsiTwo) $\vec{\lambda}$ parameters are determined in
several \pt\ bins in the range 14--70\GeV (14--50\GeV)
and in two (three) absolute rapidity bins.
Such a double-differential analysis is important to avoid obtaining diluted
results from integrating over events characterized by significantly different
kinematics~\cite{bib:Faccioli-EPJC}.

The results correspond to the polarizations of the prompt \PsiN states.
The nonprompt component, mostly from decays of B mesons, is explicitly removed
by using a proper-lifetime measurement.
A significant fraction of the \PsiOne prompt cross section is caused by feed-down
decays from the \PsiTwo (more than 8\%, increasing with \pt) and from the $\chi_c$
(more than 25\%)~\cite{bib:Faccioli-feeddown}. There
are no feed-down decays from heavier charmonium states to the \PsiTwo state, making it
particularly interesting and easier to compare the measured polarization of this state
with theoretical calculations.
The polarization extraction method uses
the dimuon invariant-mass distribution to separate the \PsiN signal
contributions from the continuum muon pairs from other processes (mostly pairs of
muons resulting from decays of uncorrelated heavy-flavour mesons).

The two-dimensional shape of the decay angular distribution (in $\cos\vartheta$ and $\varphi$)
is used to extract the three frame-dependent anisotropy parameters
in three polarization frames,
characterized by different choices of the quantization axis in the production plane:
the centre-of-mass helicity (HX) frame, where the $z$ axis coincides with the direction of the \PsiN
momentum in the laboratory;
the Collins--Soper (CS) frame~\cite{bib:CS}, whose $z$ axis is the bisector of the two
beam directions in the \PsiN\ rest frame; and the perpendicular helicity (PX) frame~\cite{Braaten:2008mz},
with the $z$ axis orthogonal to that in the CS frame.
The $y$ axis is taken, in all cases,
to be in the direction of the vector product
of the two beam directions in the charmonium rest frame,
$\vec{P}_1 \times \vec{P}_2$ and $\vec{P}_2 \times \vec{P}_1$
for positive and negative dimuon rapidities, respectively.
More details regarding these frames are provided in Ref.~\cite{bib:Faccioli-EPJC}.
The parameter \ltilde, introduced in Ref.~\cite{bib:Faccioli-PRD-FrameInv} to provide an
alternative and frame-independent characterization of the quarkonium polarization properties,
is measured simultaneously with the other parameters.
This multidimensional
approach reduces and keeps under control the smearing effects of the (unavoidable)
partial averaging of the results over the range of the production and decay kinematics.
This is important to
minimize the possible interpretation ambiguities in the comparison with theoretical
predictions and other experimental measurements~\cite{bib:Faccioli-EPJC}.

\section{CMS detector and data processing}

The CMS apparatus~\cite{Chatrchyan:2008zzk} was designed around a central element:
a superconducting solenoid of 6\unit{m} internal diameter, providing a 3.8\unit{T} field.
Within the solenoid volume are a silicon pixel and strip tracker, a lead tungstate crystal
electromagnetic calorimeter, and a brass/scintillator hadron calorimeter. Muons are measured
in gas-ionization detectors embedded in the steel return yoke outside the solenoid and made using
three technologies: drift tubes, cathode strip chambers, and resistive plate chambers.
Extensive forward calorimetry complements the coverage provided by the barrel and endcap detectors.
The main subdetectors used in this analysis are the silicon tracker and the muon system, which
enable the measurement of muon momenta over the pseudorapidity range $\abs{\eta} < 2.4$.

The events were collected using a two-level trigger system.
The first level consists of custom hardware processors and uses information from the muon
system to select events with two muons. The ``high-level trigger'' significantly reduces the
number of events written to permanent storage by requiring an opposite-sign muon pair that
fulfills certain kinematic conditions: invariant mass $2.8<M<3.35$\GeV, $\pt > 9.9$\GeV,
and $\abs{y} < 1.25$ for the \PsiOne trigger; $3.35<M<4.05$\GeV and $\pt > 6.9$\GeV
for the \PsiTwo trigger.
There is no rapidity requirement on the \PsiTwo trigger, given its lower cross section,
permitting an extra bin at forward rapidity with respect to the \PsiOne case.
No \pt\ requirement is imposed on the single muons at trigger level, only on the dimuon.
Both triggers require a dimuon vertex-fit $\chi^2$ probability greater than 0.5\%.
Events where the two muons bend towards each other in the magnetic field
are rejected to lower the trigger rate while retaining the events where
the dimuon detection efficiencies are most reliable.

The dimuons are reconstructed by combining two opposite-sign muons. The muon tracks
are required to have hits in at least 11 tracker layers, at least two of which should be in the
silicon pixel detector, and to be matched with at least one segment in the muon system.
They must have a good track-fit quality ($\chi^2$ per degree of freedom smaller than 1.8)
and point to the interaction region.
The selected muons must also be close, in pseudorapidity and azimuthal angle,
to the muon objects responsible for triggering the event.
In order to ensure accurately measured muon detection efficiencies, the analysis is
restricted to muons produced within the range $\abs{\eta} < 1.6$ and having transverse
momentum above 4.5, 3.5, and 3.0\GeV for $\abs{\eta} < 1.2$, $1.2 < \abs{\eta} < 1.4$, and
$1.4 < \abs{\eta} < 1.6$, respectively.
The continuum background due to pairs of uncorrelated muons is reduced by requiring a
dimuon vertex-fit $\chi^2$ probability larger than 1\%.
After applying all event selection criteria and background removal,
the total numbers of prompt plus nonprompt \PsiOne events are 2.3\,M and 2.4\,M
in the rapidity bins $\abs{y}<0.6$ and $0.6<\abs{y}<1.2$, respectively. The corresponding
\PsiTwo yields are 126\,k, 136\,k, and 55\,k for $\abs{y}<0.6$,
$0.6<\abs{y}<1.2$, and $1.2<\abs{y}<1.5$, respectively.
In each of these $\abs{y}$ ranges, the analysis is performed in several
\pt bins, with boundaries at
14, 16, 18, 20, 22, 25, 30, 35, 40, 50, and 70\GeV for the \PsiOne,
and 14, 18, 22, 30, and 50\GeV for the \PsiTwo.

The single-muon detection efficiencies are measured by a
tag-and-probe technique~\cite{Khachatryan:2010xn}, using
event samples collected with dedicated triggers enriched in dimuons from \PsiOne decays,
where a muon is combined with a track and the pair is required to have an invariant mass
within the range 2.8--3.4\GeV.
The measurement procedure has been validated in the
fiducial region of the analysis with detailed Monte Carlo (MC) simulation studies.
The single-muon efficiencies are precisely measured and parametrized
as a function of \pt, in eight $\abs{\eta}$ bins,
to avoid biases in the angular distributions that could mimic polarization effects.
Their uncertainties, reflecting the statistical precision of the tag-and-probe
samples and possible imperfections of the parametrization, contribute to the
systematic uncertainty in the polarization measurement.
At high dimuon \pt, when the two decay muons might be emitted relatively close to each other,
the dimuon trigger has a lower efficiency than the simple product of the two single-muon
efficiencies. Detailed MC simulations, validated with data collected with single-muon and dimuon triggers,
are used to correct these trigger-induced muon-pair correlations.

\section{Extraction of the polarization parameters}

For each \PsiN $(\pt, \abs{y})$ bin, the dimuon invariant-mass distribution is fitted,
using an unbinned maximum-likelihood fit, with an exponential function representing the
underlying continuum background and two Crystal Ball (CB) functions~\cite{CrystalBall}
representing each peak. The two CB functions have independent widths,
$\sigma_\mathrm{CB_1}$ and $\sigma_\mathrm{CB_2}$, to accommodate the changing
dimuon invariant-mass resolution within the rapidity cells, but share the same mean
$\mu_\mathrm{CB}$ and tail factors $\alpha_\mathrm{CB}$ and $n_\mathrm{CB}$ (the latter fixed to 2.5).

\begin{figure}[t]
\centering
\includegraphics[width=0.45\textwidth]{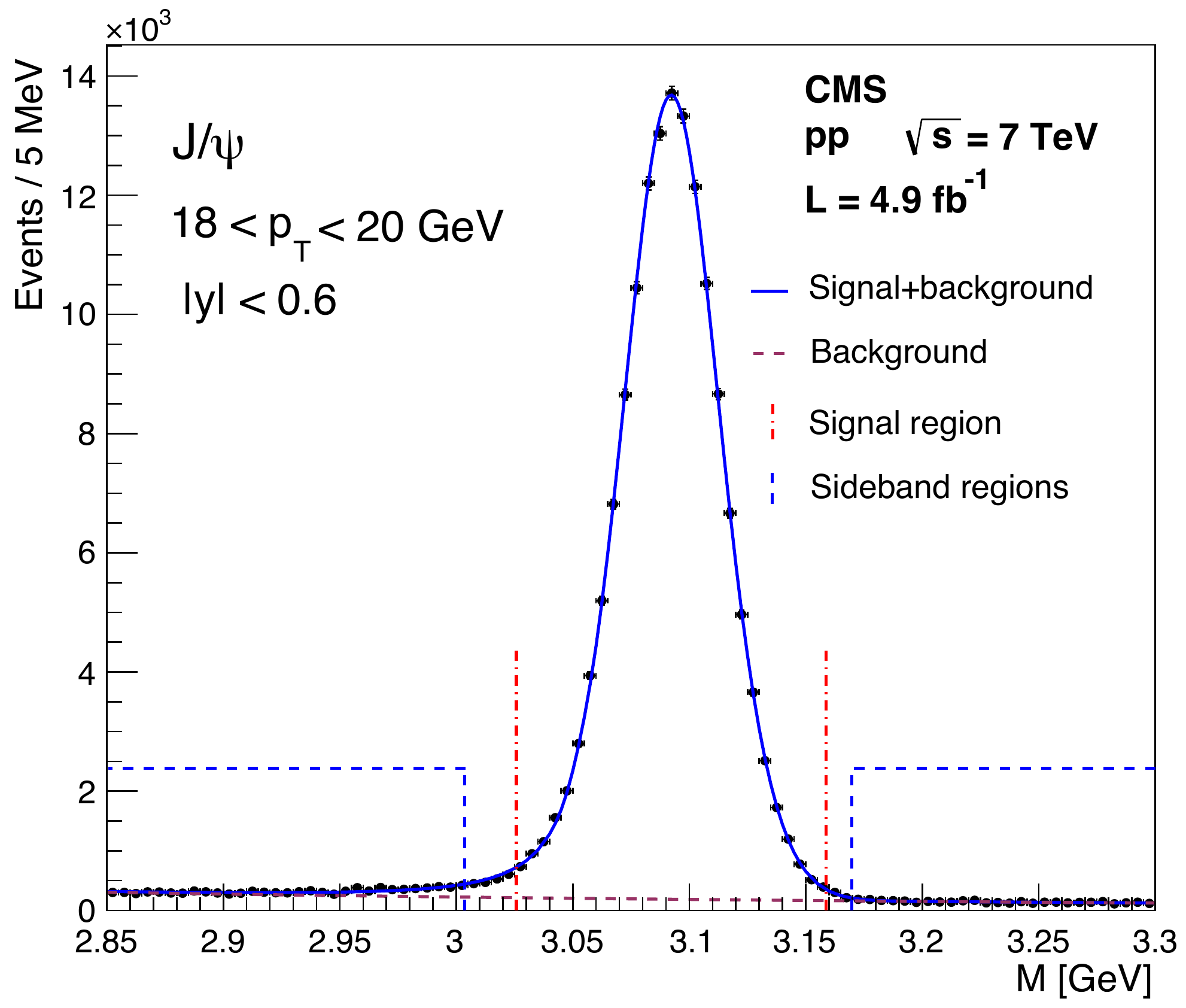}
\includegraphics[width=0.45\textwidth]{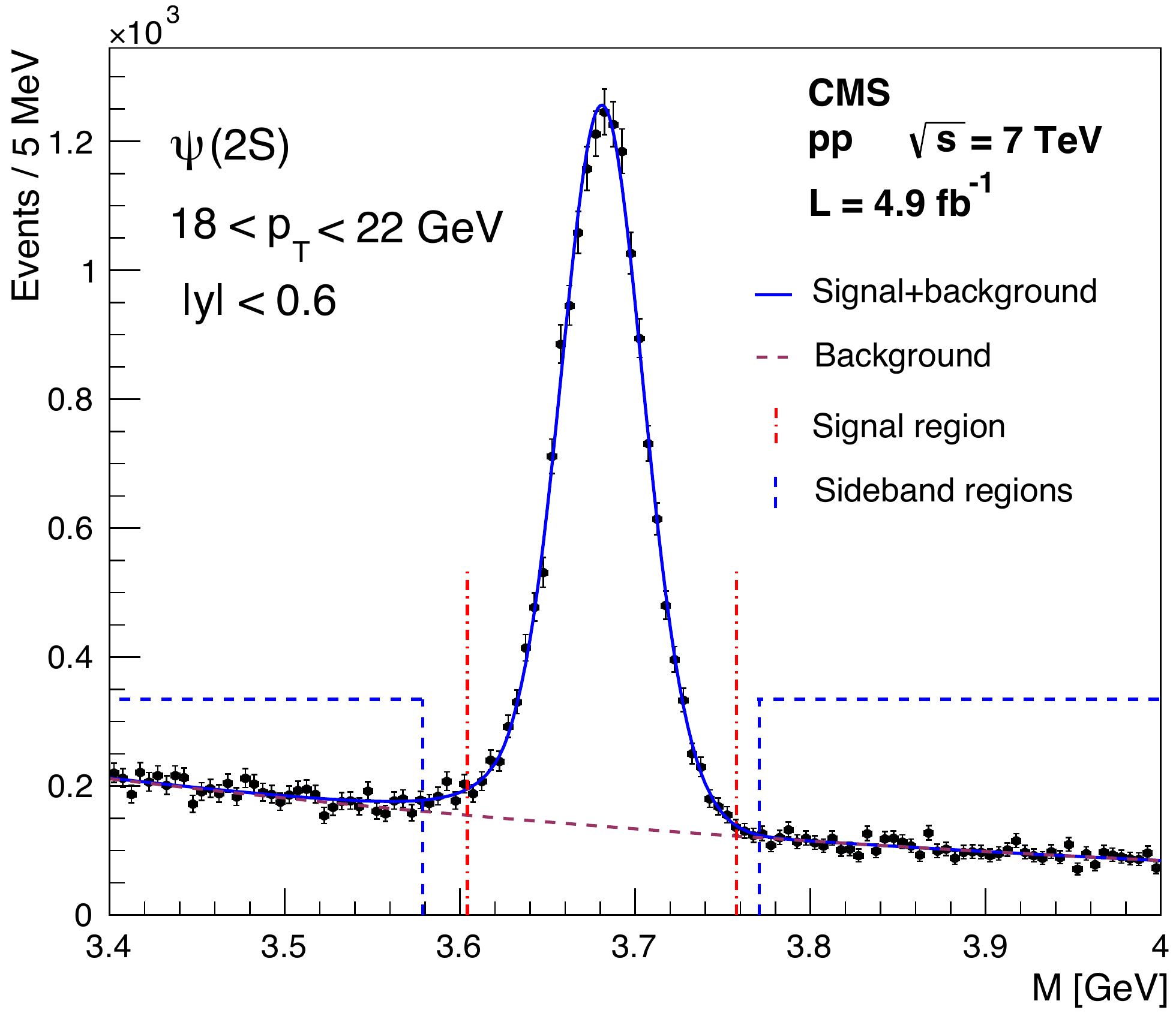}
\caption{Dimuon invariant-mass distribution in the \PsiOne (\cmsLeft) and
\PsiTwo (\cmsRight) regions for an intermediate \pt\ bin
and $\abs{y}<0.6$.
The vertical lines delimit the signal region (dot-dashed) and the mass
sidebands (dashed).
The results of the fits are shown by the solid (signal+background) and
dashed (background only) curves.}
\label{fig:dimuon_mass}
\end{figure}

Figure~\ref{fig:dimuon_mass} shows two representative dimuon invariant-mass distributions
in specific kinematic bins of the analysis.
The dimuon invariant-mass resolution $\sigma$ at the \PsiN masses is evaluated from the
fitted signal shapes, as
$\sqrt{\smash[b]{f_{\mathrm{CB}_1} \sigma_{\mathrm{CB}_1}^2 + (1-f_{\mathrm{CB}_1})\, \sigma_{\mathrm{CB}_2}^2}}$,
where $f_{\mathrm{CB}_1}$ is the relative weight of the $\mathrm{CB}_1$ function.
The \pt-integrated values are
$\sigma_{\PsiOne}=21$ and 32\MeV for $\abs{y}<0.6$ and $0.6<\abs{y}<1.2$, respectively, and
$\sigma_{\PsiN}=25$, 37, and 48\MeV for $\abs{y}<0.6$, $0.6<\abs{y}<1.2$, and
$1.2<\abs{y}<1.5$, respectively.
For each $(\pt, \abs{y})$ bin, the measured mass resolution is used
to define a $\pm 3 \sigma$ signal window around the
resonance mass~\cite{bib:PDG2012}, $m$, as well as two mass
sidebands, at lower and higher masses:
from 2.85\GeV to $m_{\PsiOne} - 4 \sigma_{\PsiOne}$ and
from $m_{\PsiOne} + 3.5 \sigma_{\PsiOne}$ to 3.3\GeV
for the \PsiOne;
from 3.4\GeV to $m_{\PsiTwo} - 4 \sigma_{\PsiTwo}$ and
from $m_{\PsiTwo} + 3.5 \sigma_{\PsiTwo}$ to 4\GeV
for the \PsiTwo.
The larger gap in the low-mass sideband definition
compared to the high-mass sideband minimizes the signal
contamination induced by the low-mass tail of the signal peaks.
The result of the invariant-mass fit provides the fraction of
continuum-background events.

To minimize the fraction of charmonia from B decays in the
sample used for the polarization measurement,
a ``prompt-signal region" is defined using the dimuon
pseudo-proper lifetime~\cite{bib:BPH-10-002},
$\ell = L_{xy} \cdot m_{\PsiN} / \pt$,
where $L_{xy}$ is the
transverse decay length in the
laboratory frame.
The measurement of $L_{xy}$ is performed after removing
the two muon tracks from the calculation of the primary vertex position;
in the case of events
with multiple collision vertices (pileup), we select the one closest to the
direction of the dimuon momentum, extrapolated towards the beam line.

\begin{figure}[htb]
\centering
\includegraphics[width=0.45\textwidth]{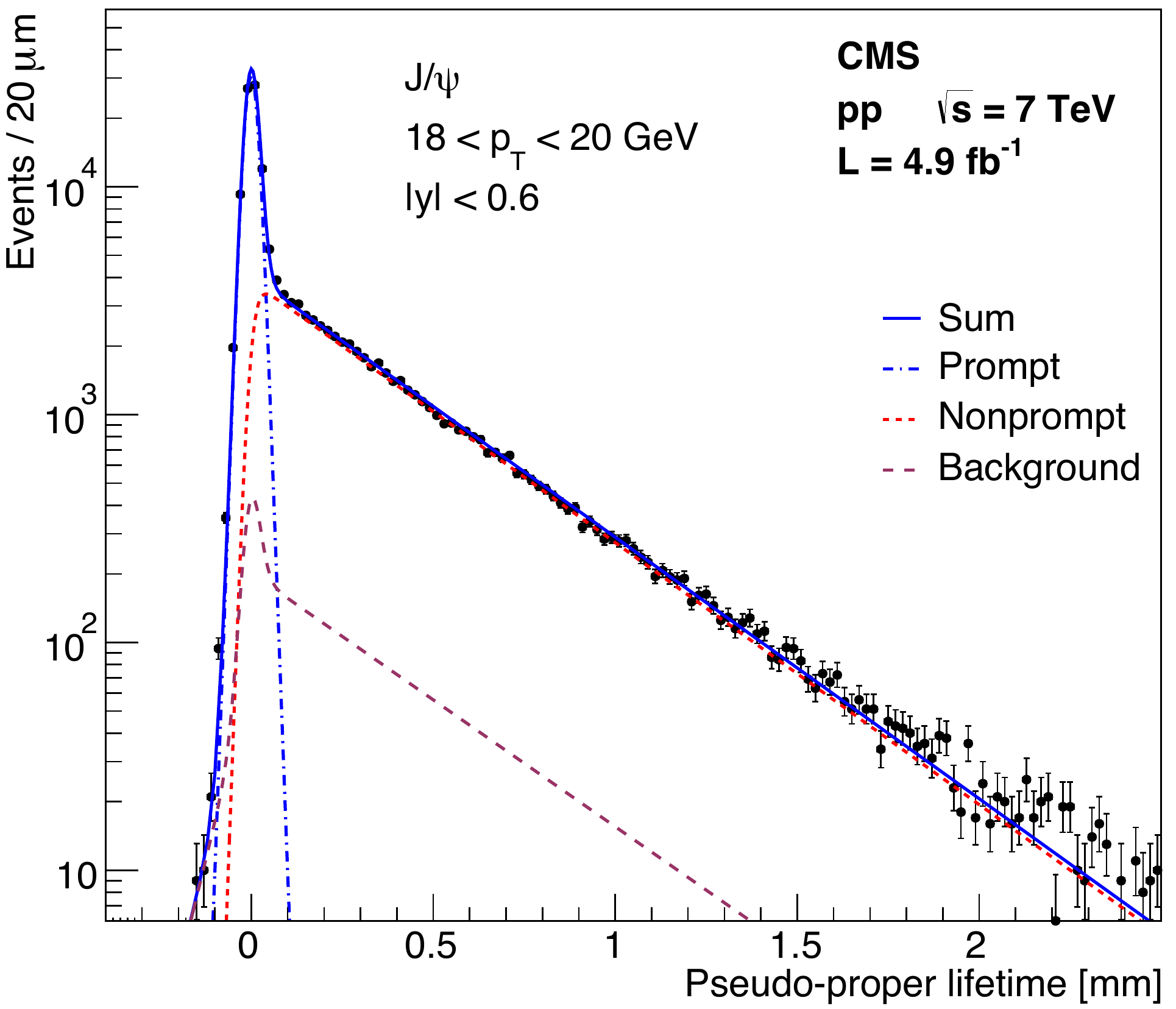}
\includegraphics[width=0.45\textwidth]{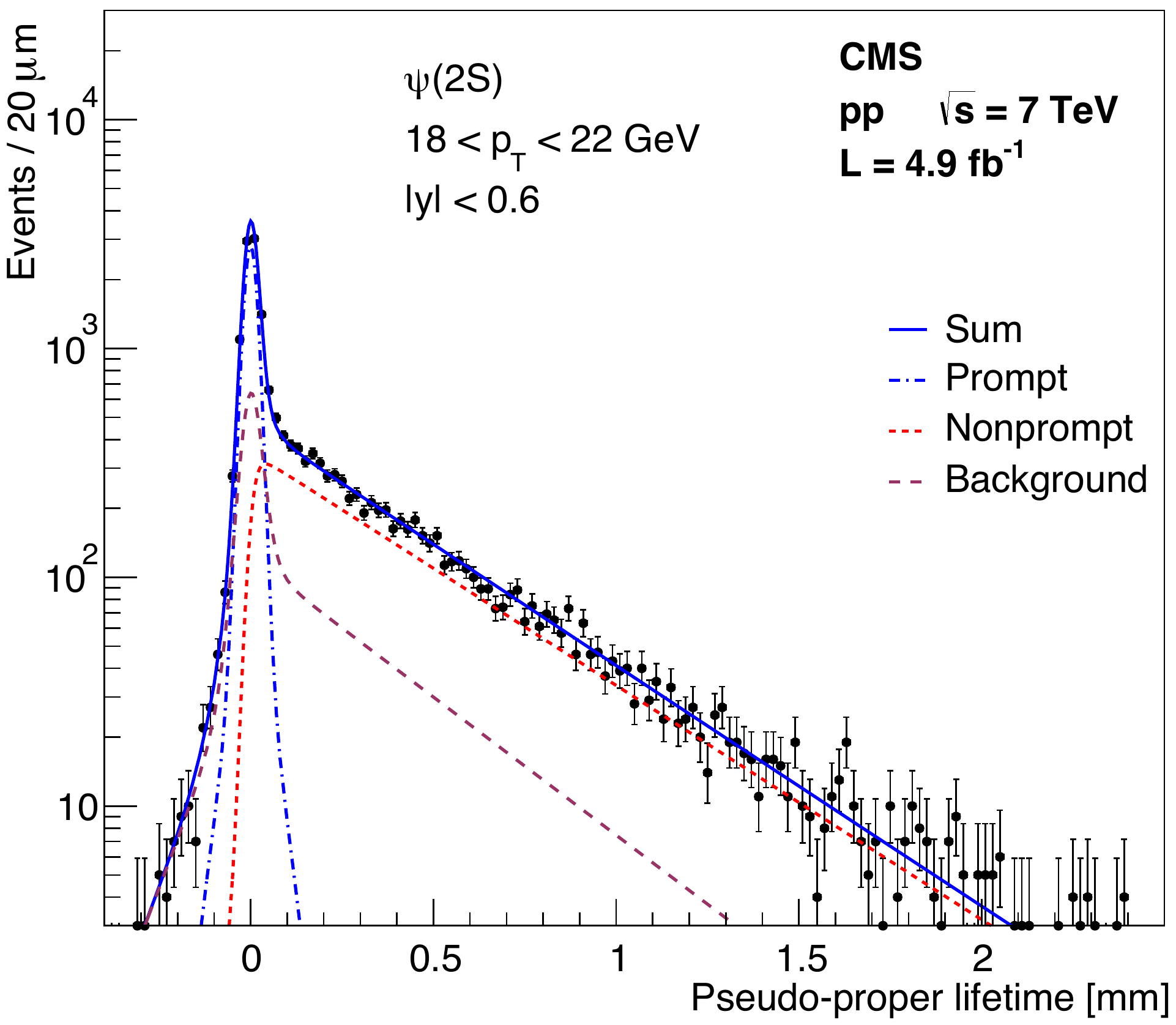}
\caption{Pseudo-proper-lifetime distribution in the \PsiOne (\cmsLeft) and
\PsiTwo (\cmsRight) mass regions for intermediate \pt\ bins and $\abs{y}<0.6$.
The results of the fits are shown by the solid curve, representing the
sum of three contributions: prompt (dash-dotted), nonprompt
(dotted), and background (dashed).}
\label{fig:lifetime}
\end{figure}

The modelling of the resolution of the pseudo-proper lifetime
exploits the per-event uncertainty information provided by the vertex reconstruction algorithm.
The prompt-signal component is modelled by the resolution function, the nonprompt
component by an exponential decay function convolved with the resolution function,
and the continuum-background component by the sum of three exponential functions,
also convolved with the resolution function.
This composite model describes the data well with a relatively small number of free
parameters.
The systematic uncertainties induced by the lifetime fit in the
polarization measurement are negligible.
Figure~\ref{fig:lifetime} shows representative pseudo-proper-lifetime
distributions  for dimuons in the two \PsiN signal regions,
together with the results of unbinned
maximum-likelihood fits, performed simultaneously in the signal region
and mass sidebands.

\begin{figure}[htb]
\centering
\includegraphics[width=\cmsFigWidth]{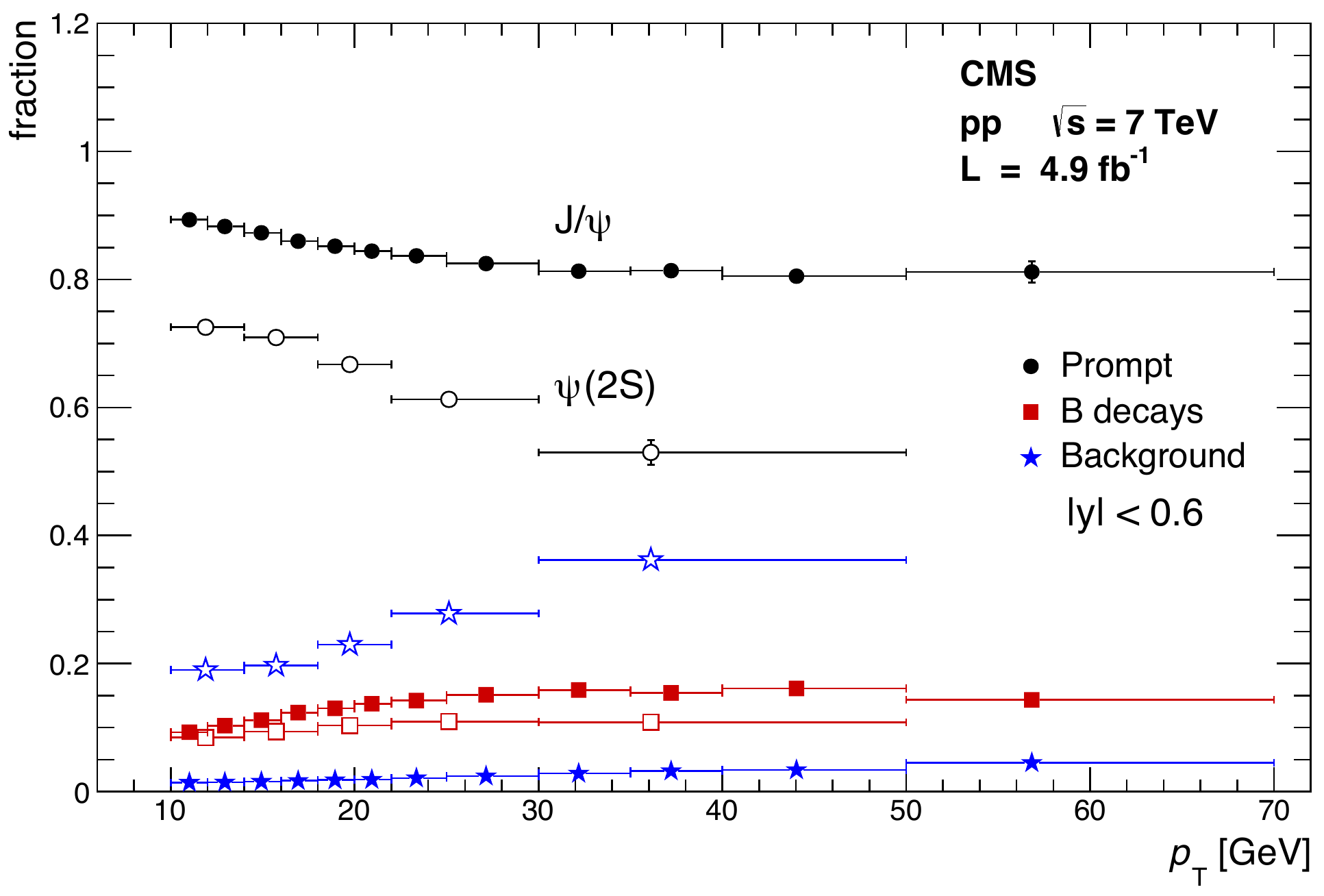}
\caption{Fractions of prompt charmonium (circles),
charmonium from B decays (squares), and
continuum-background (stars) events in the prompt-signal mass-lifetime
\PsiOne (closed symbols) and \PsiTwo (open symbols) regions versus
the dimuon \pt for $\abs{y}<0.6$.
A sideband subtraction technique removes the B and continuum
backgrounds from the polarization analysis.}
\label{fig:fractions}
\end{figure}

The prompt-signal regions, dominated by prompt charmonium events,
are defined as ${\pm}3 \sigma_{\ell}$ signal windows around $\ell=0$,
where the lifetime resolution, $\sigma_{\ell}$, is measured to be
(for the phase space probed in this analysis)
in the range 12--25\micron, improving with increasing dimuon \pt.
The fractions of charmonia from B decays (\fNP) and
continuum-background events (\fbg) included in these regions
are shown in Fig.~\ref{fig:fractions} versus
the dimuon \pt, for $\abs{y}<0.6$.

For each \PsiN state, the angular distribution of the continuum background
is modelled as the weighted sum of the distributions measured in the two
mass sidebands (restricted to the prompt-lifetime region),
with weights derived under the assumption that
the background distribution changes
linearly with the dimuon mass.
This assumption is validated by comparing the (small) differences of the effective
background polarizations measured in the four dimuon invariant-mass sidebands.
The angular distribution of the \PsiN from B decays is modelled using the
events in the \PsiN mass peak belonging to the ``nonprompt-lifetime region",
$\ell > 3 \sigma_{\ell}$, after subtracting the corresponding continuum-background
contribution, interpolated from the nonprompt mass-sideband regions.
As a cross-check of the analysis, the polarization of the nonprompt component was also measured, in
two lifetime regions ($\ell > 3 \sigma_{\ell}$ and $\ell > 5 \sigma_{\ell}$), with consistent results.

The total background is the sum of the continuum-background and charmonia from B decays
present in the prompt-signal region.
To remove the background component, a fraction $\fbgtot = \fbg + \fNP$
of the events is randomly selected by a  procedure based on the likelihood-ratio
$L_\mathrm{B} / L_\mathrm{S+B}$, where $L_\mathrm{B}$ ($L_\mathrm{S+B}$)
is the likelihood for an event under the background-only (signal-plus-background) hypothesis.
This selection
operates in such a way that the chosen events are distributed according
to the $(\pt, \abs{y}, M, \costh, \varphi)$ distribution of the background model.
The randomly selected events are removed from the sample.

The remaining (signal-like) events are used to calculate the
posterior probability density (PPD)
of the prompt-\PsiN polarization
parameters ($\vec{\lambda}$) for each kinematic bin,
\begin{linenomath}
\begin{equation}
\label{eq:total_likelihood}
  \mathcal{P}(\vec{\lambda})= \prod_{i} \mathcal{E}(\vec{p}_1^{\,(i)},\vec{p}_2^{\,(i)}),
\end{equation}
\end{linenomath}
where $\mathcal{E}$ is the probability density as a function of the two muon
momenta $\vec{p}_{1,2}$ in event $i$.
Uniform priors are used in the full $\vec{\lambda}$ parameter space.
Many previous polarization measurements were dependent on
assumptions made about the production kinematics because of the use of simulated acceptance
and efficiency dilepton $(\costh,\varphi)$ maps, averaged over all events in the
considered kinematic cell.
This analysis, instead, uses the efficiencies measured as a function of muon momentum,
attributing to each event a probability dependent on the full event kinematics
(not only on $\costh$ and $\varphi$) and on the values of the polarization parameters.
The event probability is calculated as
\begin{linenomath}
\begin{equation}
\label{eq:event_probability_definition}
\mathcal{E}(\vec{p}_1,\vec{p}_2)= \frac{1}{\mathcal{N}(\vec{\lambda})} \,
W(\costh,\varphi|\vec{\lambda}) \, \epsilon(\vec{p}_1,\vec{p}_2),
\end{equation}
\end{linenomath}
where $W$ is defined in Eq.~(\ref{eq:angular_distribution}) and
$\epsilon(\vec{p}_1,\vec{p}_2)$ is the dimuon detection efficiency.
The $\mathcal{N}(\vec{\lambda})$ normalization factor is obtained from integrating
$W \, \epsilon$ over $\costh$ and $\varphi$,
\begin{equation}
\label{eq:normalization}
\begin{split}
\mathcal{N}=
\frac{1}{(3 + \lambda_{\vartheta})}&
\bigg[
\left( \iint \epsilon(\vec{p}_1,\vec{p}_2) \; \rd\costh\, \rd\varphi\right) +\\
+\lambda_{\vartheta}&
\left( \iint \cos^2 \vartheta \; \epsilon(\vec{p}_1,\vec{p}_2) \; \rd\cos\vartheta\, \rd\varphi \right)\, + \\
+\lambda_{\varphi}&
\left( \iint \sin^2 \vartheta \cos 2 \varphi \; \epsilon(\vec{p}_1,\vec{p}_2) \; \rd\cos\vartheta\, \rd\varphi \right)\, +\\
+\lambda_{\vartheta \varphi}&
\left( \iint \sin 2 \vartheta \cos \varphi \; \epsilon(\vec{p}_1,\vec{p}_2) \; \rd\cos\vartheta\, \rd\varphi \right)
\bigg].
\end{split}
\end{equation}
To perform this integration, $\epsilon(\vec{p}_1,\vec{p}_2)$ is expressed in terms of
$\costh$ and $\varphi$ using the background-removed $(\pt, \abs{y}, M)$ distributions.
The background-removal procedure is repeated 50 times to minimize the statistical
fluctuations associated with its random nature, and the PPD
is obtained as the average of the 50 individual densities.
The value 50 is very conservative; 20 iterations would have been sufficient to
provide stable results.

\begin{figure}[htb]
\centering
\includegraphics[width= 0.42\textwidth]{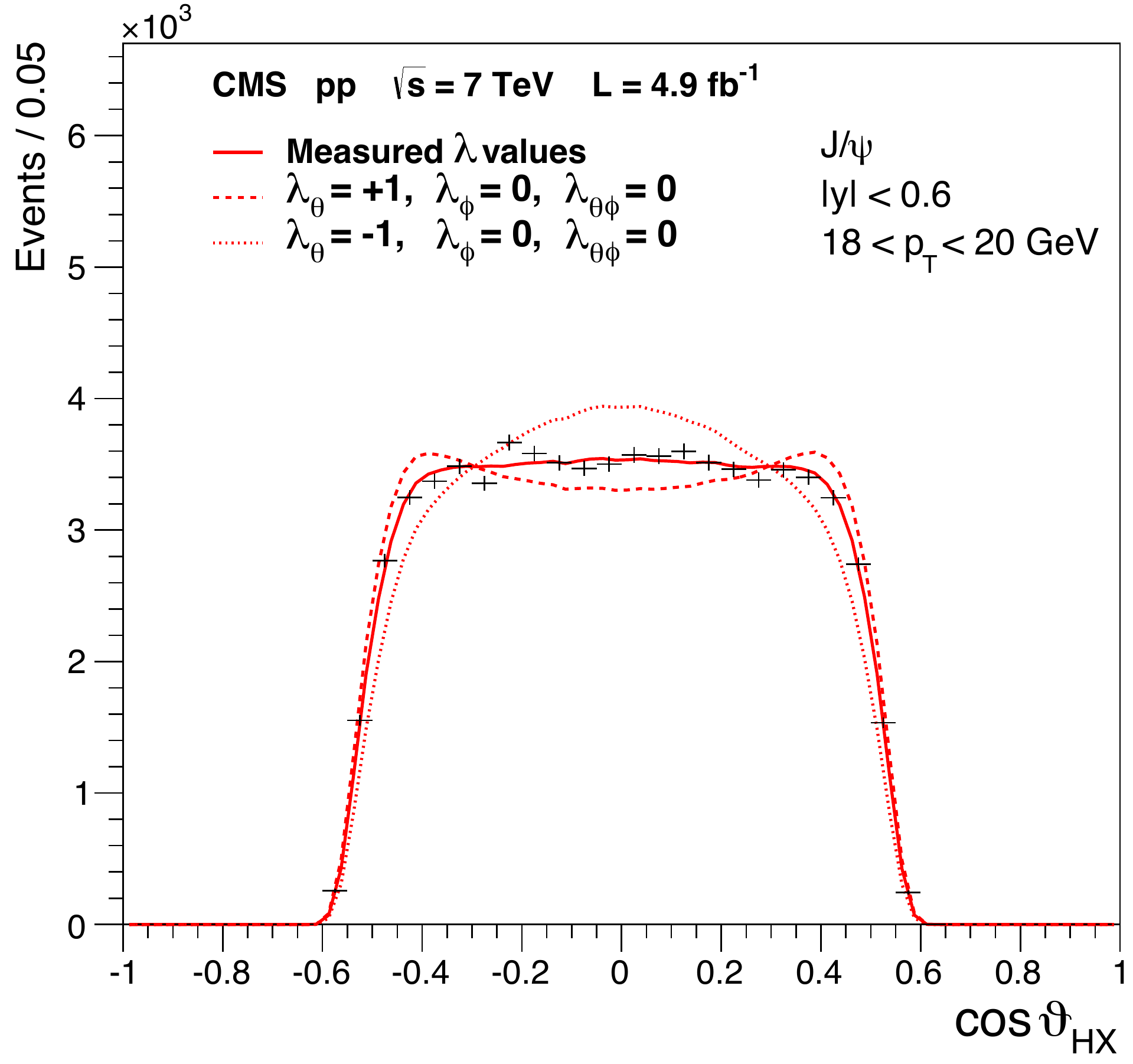}
\includegraphics[width= 0.42\textwidth]{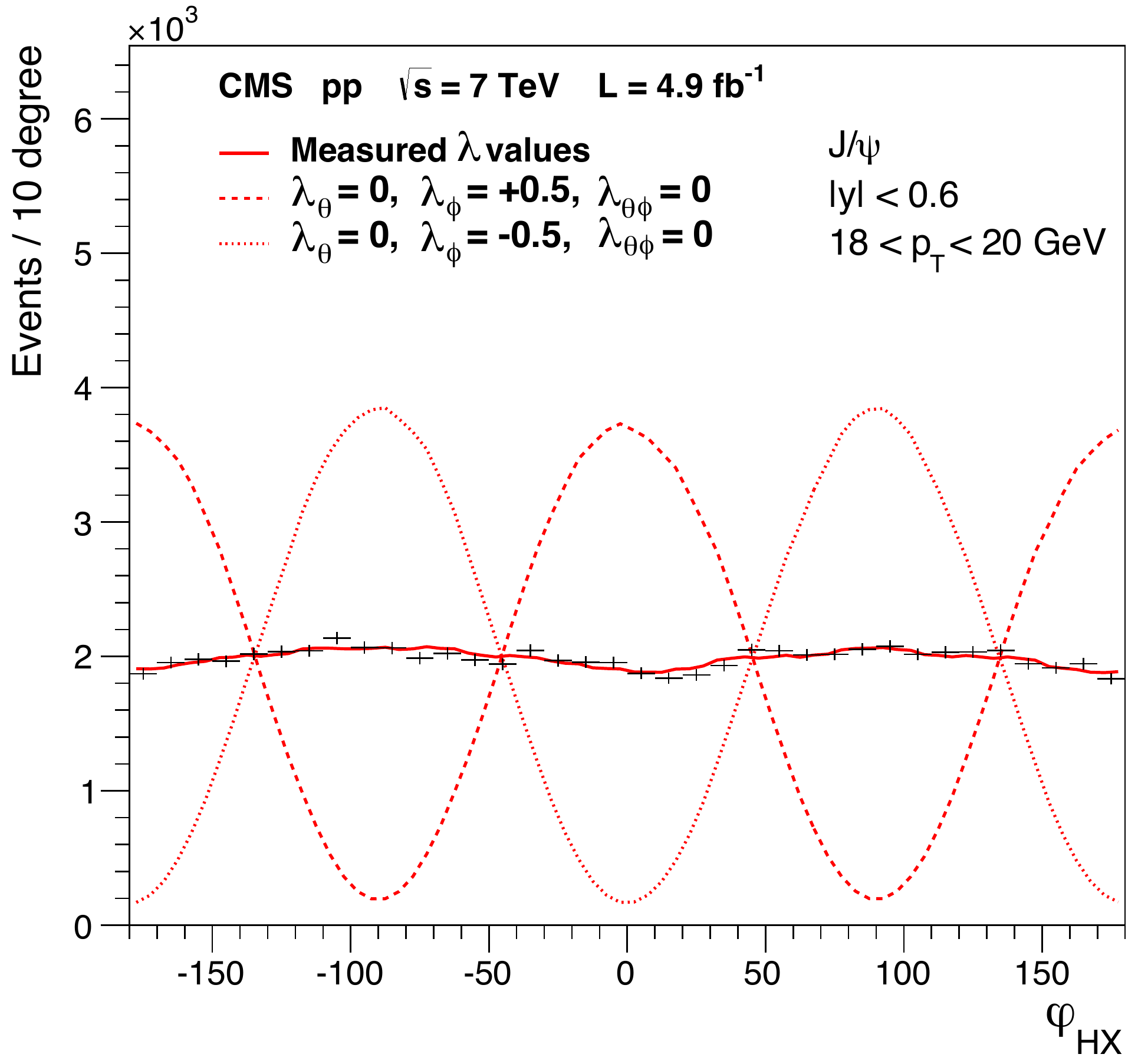}
\caption{Frequency distributions of $\costh$ (\cmsLeft) and $\varphi$ (\cmsRight) angular
variables, in the HX frame for the \PsiOne in an intermediate \pt\ bin and $\abs{y}<0.6$.
The curves represent the expected distributions for two extreme polarization scenarios
(dashed and dotted lines defined in the legends) and for the measured $\vec{\lambda}$
(solid lines).}
\label{fig:angularProjections}
\end{figure}

Figure~\ref{fig:angularProjections} illustrates the measured $\costh$ and $\varphi$
distributions in the HX frame for the case of \PsiOne signal events in the kinematic bin $\abs{y}<0.6$ and
$18<\pt<20$\GeV, after background removal.
The data points are compared to curves reflecting the ``best fit"
(solid lines) as well as two extreme scenarios (dashed and dotted lines), corresponding
to the \lth, \lph, and \ltp\ values reported in the legends of the plots.

Most of the systematic uncertainties we have considered were studied and quantified
(for each charmonium and each kinematic bin) with pseudo-experiments
based on simulated events.
Each test evaluates a specific systematic uncertainty and uses 50 statistically independent
event samples, individually generated and reconstructed. The difference between the
median of the 50 obtained polarization parameters and the injected values provides the
systematic uncertainty corresponding to the effect under study.
In particular, several signal and background polarization scenarios
have been used to evaluate the reliability of the analysis framework,
including extreme signal polarizations in the highest-\pt bins of the analysis,
where the dimuon trigger inefficiency has the strongest effect.
Possible residual biases in the muon or dimuon efficiencies, resulting from the tag-and-probe
measurement precision or from the efficiency parametrization, could affect the extraction of the
polarization parameters. This effect is evaluated by applying uncertainty-based changes to the
used efficiencies.
The systematic uncertainty resulting from the unknown background angular distribution
under the signal peak is evaluated using the measured data, by changing the relative weights
of the low- and high-mass sidebands in the background model between 0.25 and 0.75,
very different from the measured values of $\approx$0.5.
The resulting uncertainty is negligible, as expected given the small magnitude of the
background and the proximity of the mass sidebands to the charmonia peaks.
The systematic uncertainty associated with the definition of the prompt-signal region is
evaluated as the difference between the MC simulation results obtained with a $\pm 3 \sigma_{\ell}$
window and with no pseudo-proper-lifetime requirement.

The \PsiTwo polarization uncertainties are dominated by statistics limitations
in all $(\pt, \abs{y})$ bins. In the \PsiOne case, at high \pt\ the uncertainties are dominated
by the statistical accuracy,
while for $\pt \lesssim 30$\GeV they are determined by systematic effects.
The largest among these include
the single-muon ($\approx$0.1, 0.02, and 0.03) and
dimuon ($\approx$0.05, 0.03, and 0.02) efficiencies, and
the prompt-region definition ($\approx$0.03, 0.02, and 0.01);
the values given correspond to the systematic uncertainties
for \lth, \lph, and \ltp, respectively, in the HX frame, averaged
over the rapidity bins.

\begin{figure}[h!tb]
\centering
\includegraphics[width= 0.42\textwidth]{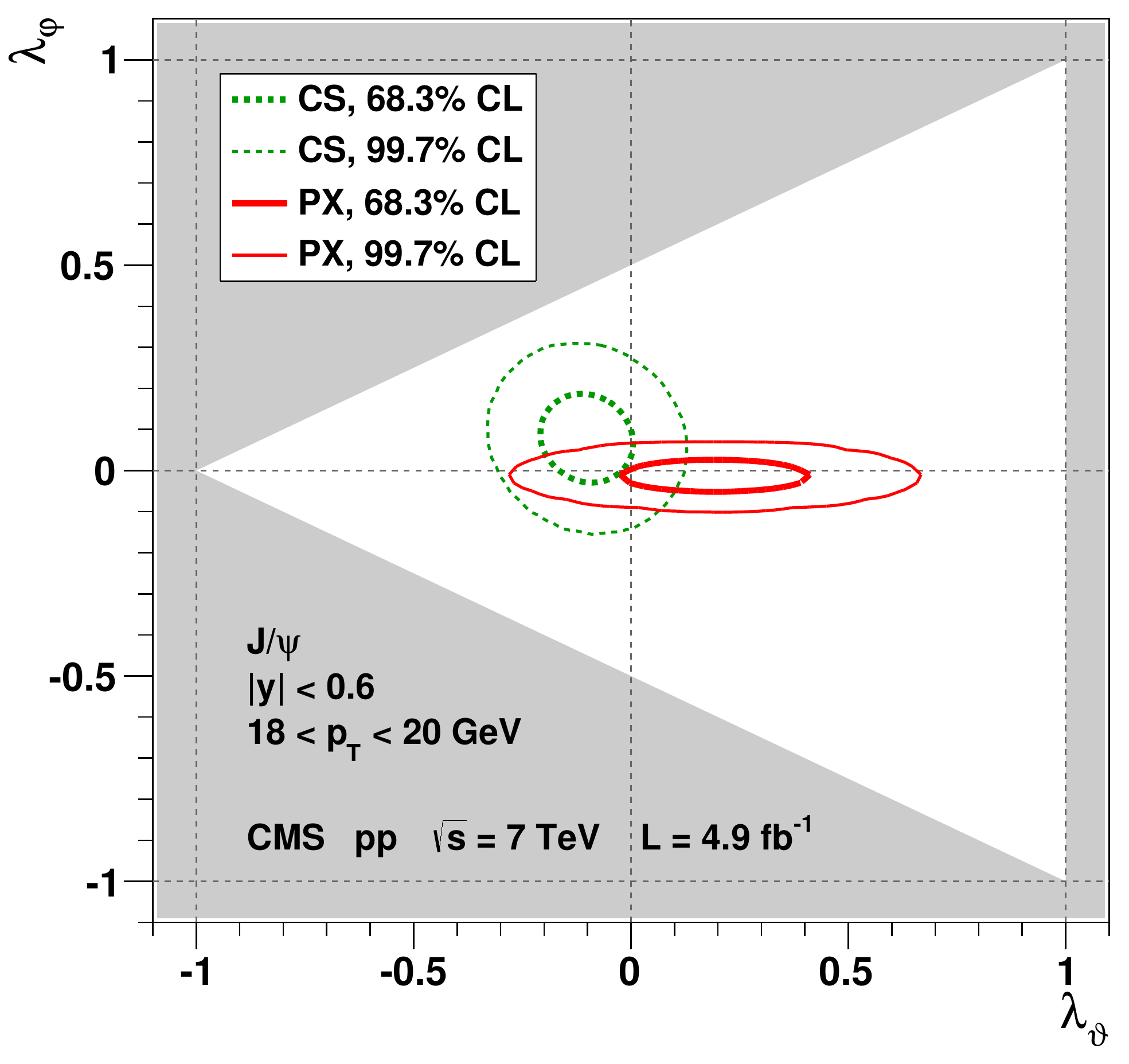}
\includegraphics[width= 0.42\textwidth]{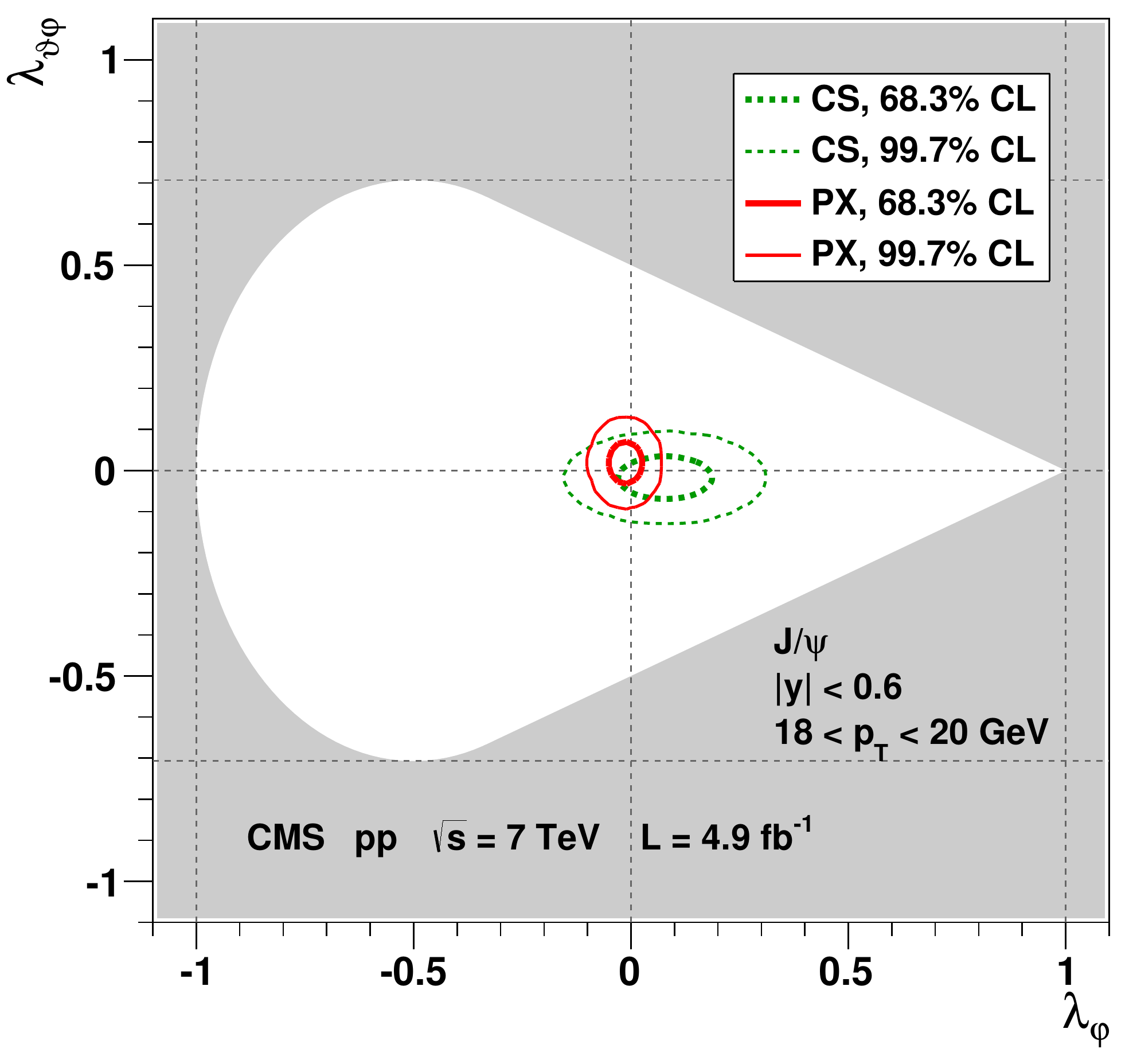}
\caption{Two-dimensional marginals
of the PPD in the \lph\ vs.\ \lth\ (\cmsLeft) and
\ltp\ vs.\ \lph\ (\cmsRight) planes, for \PsiOne with $\abs{y}<0.6$ and
$18<\pt<20$\GeV. The 68.3\% and 99.7\% CL
total uncertainties
are shown for the CS and PX frames. The shaded areas
represent physically forbidden regions of parameter
space~\cite{bib:Faccioli-shapes}.}
\label{fig:2Dcorrelations}
\end{figure}

The final PPD of the polarization parameters is the average of the PPDs corresponding
to all hypotheses considered in the determination of the systematic uncertainties.
The central value of each polarization parameter, for each kinematic bin, is evaluated as the
mode of the associated one-dimensional marginal posterior, which is calculated by
numerical integration.
The corresponding %(total)
uncertainties, at a given confidence level (CL), are given by the
$[\lambda_1,\lambda_2]$ intervals, defined such that each of the regions
$[-\infty,\lambda_1]$ and $[\lambda_2,\infty]$ integrates to half of
$(1-\mathrm{CL})$ of the marginal PPD.
Two-dimensional marginal posteriors
provide information about correlations
between the measurements of the three $\lambda$ parameters.
As an example, Fig.~\ref{fig:2Dcorrelations} shows the
two-dimensional marginals for
\lph\ vs.\ \lth\ (\cmsLeft) and \ltp\ vs.\ \lph\ (\cmsRight) measured from
\PsiOne at $\abs{y}<0.6$ and $18<\pt<20$\GeV,
displaying the 68.3\% and 99.7\% CL contours for the CS and PX frames.
The figure also indicates the physically allowed regions for the decay of a $J=1$ particle;
this region does not affect the calculation of the PPD anywhere in the analysis.
For visibility reasons, the HX curves are not shown;
in the phase space of this analysis (mid-rapidity and
relatively high \pt), the HX and PX frames are almost identical.

\section{Results}

\begin{figure*}[hbtp]
\centering
\includegraphics[width=0.85\textwidth]{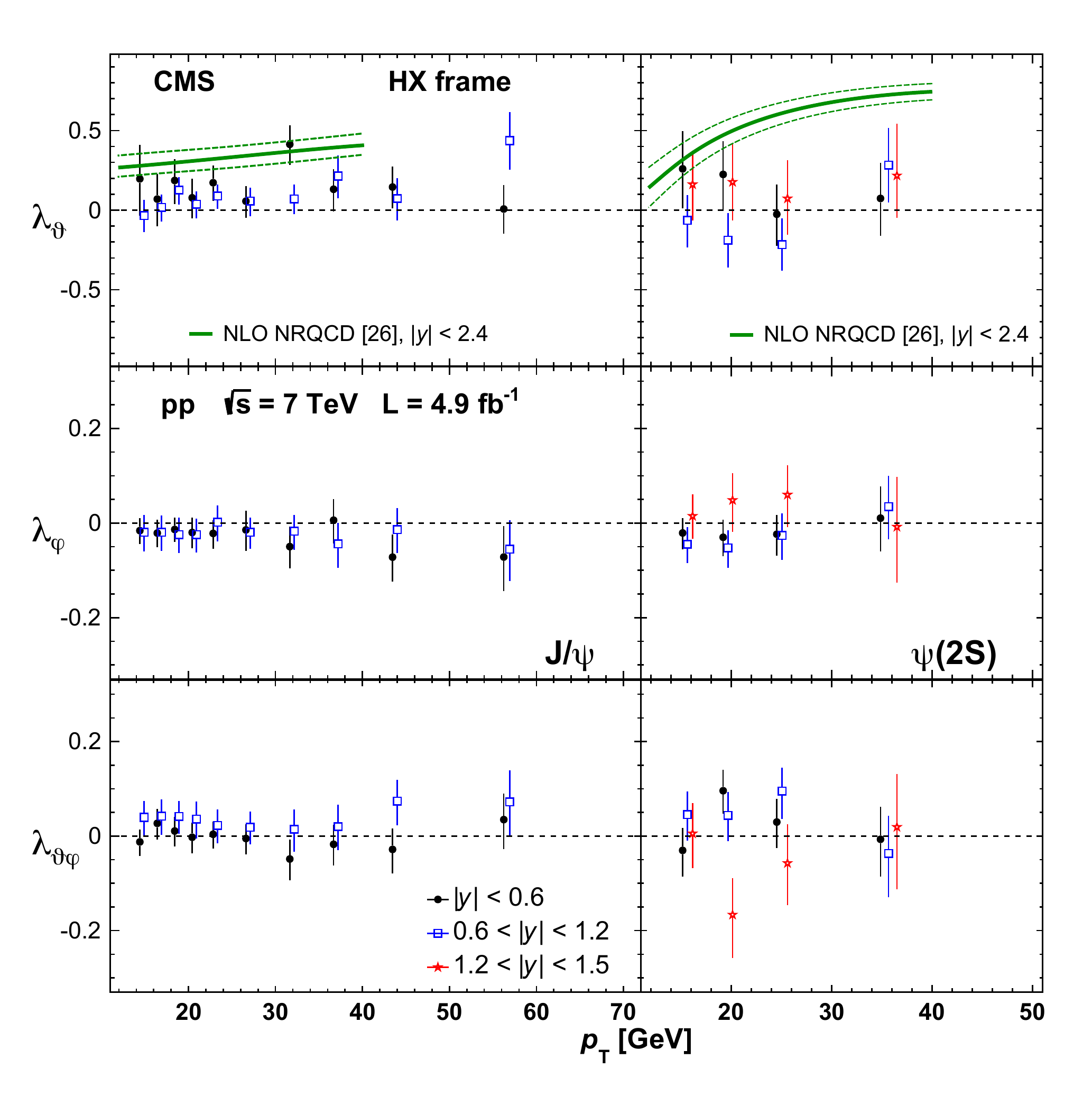}
\caption{Polarization parameters \lth, \lph, and \ltp\
measured in the HX frame for prompt \PsiOne (left) and \PsiTwo (right) mesons,
as a function of \pt\ and for several $\abs{y}$ bins.
The error bars represent total uncertainties (at 68.3\% CL).
The curves in the top two panels represent calculations of \lth\ from NLO NRQCD~\cite{Gong:2012ug},
the dashed lines illustrating their uncertainties.}
\label{fig:lambdas_HX}
\centering
\includegraphics[width=0.85\textwidth]{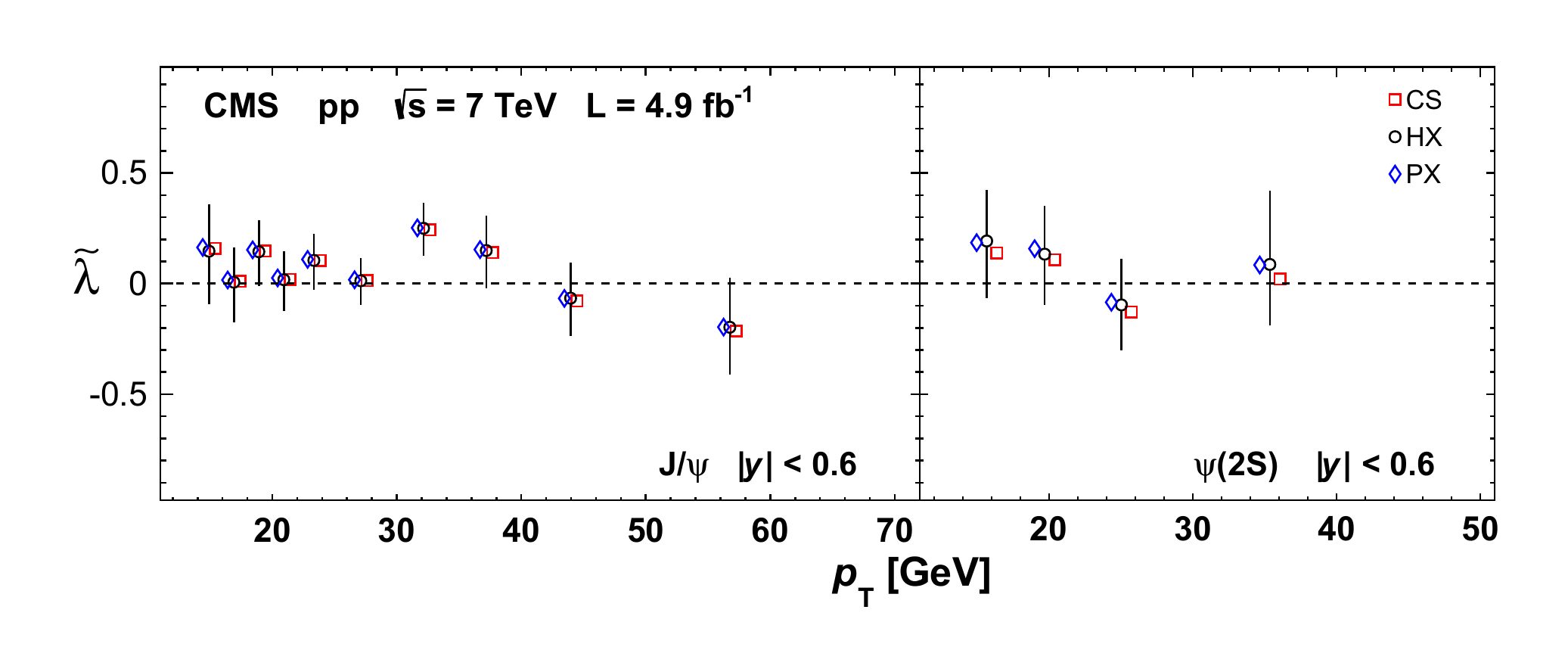}
\caption{Values of the frame-independent parameter \ltilde\ for the \PsiOne (left) and \PsiTwo (right) measured
in the CS, HX, and PX frames, as a function of \pt\ and for $\abs{y}<0.6$.
The error bars represent total uncertainties (at 68.3\% CL).}
\label{fig:lambdaTilde}
\end{figure*}

The frame-dependent $\lambda$ parameters measured in the HX frame are presented,
for both charmonia, in Fig.~\ref{fig:lambdas_HX}, as a function of \pt\ and $\abs{y}$.
The average values of \pt\ and $\abs{y}$ are given in \ifthenelse{\boolean{cms@external}}{the supplemental material}{Table~\ref{tab:MeanKinematics}
of Appendix~\ref{app:supp-mat} for each kinematic bin}.
The solid curves in the top two panels of Fig.~\ref{fig:lambdas_HX} represent next-to-leading order (NLO) NRQCD calculations~\cite{Gong:2012ug}
of the \lth\ parameter for prompt \PsiOne and \PsiTwo mesons as a function of \pt for $|y|<2.4$.
The dashed lines give an estimate of the uncertainties in the theoretical predictions.
The measured values of \lth\ are in clear disagreement with these NLO NRQCD calculations.
Figure~\ref{fig:lambdaTilde} displays the frame-invariant parameter, \ltilde,
measured in the CS, HX, and PX frames, for the rapidity range $\abs{y}<0.6$.
The three sets of \ltilde\ measurements are in good agreement, as required
in the absence of unaddressed systematic effects; the same consistency is
also observed in the other rapidity bins.
All the results for \lth, \lph, \ltp, and \ltilde, for the two \PsiN states and in the three frames
considered in this analysis, including the total 68.3\%, 95.5\%, and 99.7\% CL uncertainties
and the 68.3\% CL statistical uncertainties, are tabulated in \ifthenelse{\boolean{cms@external}}{the supplemental material}{Appendix~\ref{app:supp-mat}}.

None of the three polarization frames shows large polarizations,
excluding the possibility that a significant polarization could remain undetected
because of smearing effects induced by inappropriate frame choices~\cite{bib:Faccioli-EPJC}.
While a small prompt \PsiOne polarization can be interpreted as reflecting a
mixture of directly produced mesons with those produced in the decays of
heavier (P-wave) charmonium states, this explanation cannot
apply to the \PsiTwo state, unaffected by feed-down decays from
heavier charmonia.

\section{Summary}

In summary, the polarizations of prompt \PsiOne and \PsiTwo mesons produced in
\Pp\Pp\ collisions at $\sqrt{s} = 7$\TeV have been determined
as a function of the \PsiN \pt in two or three rapidity ranges,
extending well beyond the domains probed by previous experiments,
and in three different polarization frames, using both frame-dependent
and frame-independent parameters.
All the measured $\lambda$ parameters are close to zero,
excluding large polarizations in the explored kinematic regions.
These results are in clear disagreement with existing NLO NRQCD
calculations~\cite{Butenschoen:2012px,Gong:2012ug,Chao:2012iv}
and provide a good basis for significant improvements in the
understanding of quarkonium production in high-energy hadron collisions.
\section*{Acknowledgements}

We congratulate our colleagues in the CERN accelerator departments for the excellent performance of the LHC and thank the technical and administrative staffs at CERN and at other CMS institutes for their contributions to the success of the CMS effort. In addition, we gratefully acknowledge the computing centres and personnel of the Worldwide LHC Computing Grid for delivering so effectively the computing infrastructure essential to our analyses. Finally, we acknowledge the enduring support for the construction and operation of the LHC and the CMS detector provided by the following funding agencies: BMWF and FWF (Austria); FNRS and FWO (Belgium); CNPq, CAPES, FAPERJ, and FAPESP (Brazil); MEYS (Bulgaria); CERN; CAS, MoST, and NSFC (China); COLCIENCIAS (Colombia); MSES (Croatia); RPF (Cyprus); MoER, SF0690030s09 and ERDF (Estonia); Academy of Finland, MEC, and HIP (Finland); CEA and CNRS/IN2P3 (France); BMBF, DFG, and HGF (Germany); GSRT (Greece); OTKA and NKTH (Hungary); DAE and DST (India); IPM (Iran); SFI (Ireland); INFN (Italy); NRF and WCU (Republic of Korea); LAS (Lithuania); CINVESTAV, CONACYT, SEP, and UASLP-FAI (Mexico); MSI (New Zealand); PAEC (Pakistan); MSHE and NSC (Poland); FCT (Portugal); JINR (Armenia, Belarus, Georgia, Ukraine, Uzbekistan); MON, RosAtom, RAS and RFBR (Russia); MSTD (Serbia); SEIDI and CPAN (Spain); Swiss Funding Agencies (Switzerland); NSC (Taipei); ThEPCenter, IPST and NSTDA (Thailand); TUBITAK and TAEK (Turkey); NASU (Ukraine); STFC (United Kingdom); DOE and NSF (USA).
Individuals have received support from the Marie-Curie programme and the European Research Council and EPLANET (European Union); the Leventis Foundation; the A.P.\ Sloan Foundation; the Alexander von Humboldt Foundation; the Belgian Federal Science Policy Office; the Fonds pour la Formation \`a la Recherche dans l'Industrie et dans l'Agriculture (FRIA-Belgium); the Agentschap voor Innovatie door Wetenschap en Technologie (IWT-Belgium); the Ministry of Education, Youth and Sports (MEYS) of Czech Republic; the Council of Science and Industrial Research, India; the Compagnia di San Paolo (Torino); the HOMING PLUS programme of Foundation for Polish Science, cofinanced by EU, Regional Development Fund; and the Thalis and Aristeia programmes cofinanced by EU-ESF and the Greek NSRF.

\bibliography{auto_generated}   % will be created by the tdr script.

\ifthenelse{\boolean{cms@external}}{}{
\vfill\newpage

\appendix
\section{Supplemental material\label{app:supp-mat}}

\input{supp-mat.tex}

}
\cleardoublepage \section{The CMS Collaboration \label{app:collab}}\begin{sloppypar}\hyphenpenalty=5000\widowpenalty=500\clubpenalty=5000\input{BPH-13-003-authorlist.tex}\end{sloppypar}
\end{document}

%% file: supp-mat.tex
Table~\ref{tab:MeanKinematics} gives the average values of \pt\ and $|y|$ for each kinematical bin.
Tables~\ref{tab:first}--\ref{tab:last} list the results of the angular anisotropy parameters \lth, \lph, and \ltp, and the frame-invariant parameter \ltilde,
for the two \PsiN\ states in the Collins-Soper (CS), helicity (HX), and perpendicular helicity (PX) frames,
along with their total uncertainties (68.3\%, 95.5\%, and 99.7\% CL) and statistical uncertainties only (68.3\% CL)
in different bins of \PsiN\ transverse momentum \pt\ and absolute rapidity $|y|$.

\begin{table}[h]
\centering
\caption{Average \pt\ (in GeV) and $|y|$, for each \PsiN\ kinematical bin.}
\begin{tabular}{ccccccc}
\hline\noalign{\smallskip}
\multirow{2}{*}{\pt\ (\GeVns)}
& \multicolumn{2}{c}{$0.0 < |y| < 0.6$ }
& \multicolumn{2}{c}{$0.6 < |y| < 1.2$ }
& \multicolumn{2}{c}{$1.2 < |y| < 1.5$ } \\ \noalign{\smallskip}
\multicolumn{1}{c}{}
& $\langle p_{\rm T} \rangle$ & $\langle |y| \rangle$
& $\langle p_{\rm T} \rangle$ & $\langle |y| \rangle$
& $\langle p_{\rm T} \rangle$ & $\langle |y| \rangle$ \\ \noalign{\smallskip}
\hline \noalign{\smallskip} \multicolumn{7}{c}{\PsiOne} \\ \noalign{\smallskip}
\hline \noalign{\smallskip}
10--12 & $11.0$ & $0.30$ & $11.0$ & $0.92$ & -- & -- \\
12--14 & $12.9$ & $0.30$ & $12.9$ & $0.91$ & -- & -- \\
14--16 & $14.9$ & $0.30$ & $14.9$ & $0.90$ & -- & -- \\
16--18 & $16.9$ & $0.30$ & $16.9$ & $0.90$ & -- & -- \\
18--20 & $18.9$ & $0.30$ & $18.9$ & $0.89$ & -- & -- \\
20--22 & $20.9$ & $0.30$ & $20.9$ & $0.89$ & -- & -- \\
22--25 & $23.3$ & $0.30$ & $23.3$ & $0.89$ & -- & -- \\
25--30 & $27.1$ & $0.30$ & $27.1$ & $0.89$ & -- & -- \\
30--35 & $32.2$ & $0.30$ & $32.2$ & $0.89$ & -- & -- \\
35--40 & $37.2$ & $0.30$ & $37.2$ & $0.88$ & -- & -- \\
40--50 & $44.0$ & $0.30$ & $44.0$ & $0.89$ & -- & -- \\
50--70 & $56.8$ & $0.30$ & $56.9$ & $0.88$ & -- & -- \\
\hline \noalign{\smallskip} \multicolumn{7}{c}{\PsiTwo} \\ \noalign{\smallskip}
\hline \noalign{\smallskip}
10--14 & $11.9$ & $0.30$ & $11.8$ & $0.92$ & $11.8$ & $1.33$ \\
14--18 & $15.7$ & $0.30$ & $15.6$ & $0.90$ & $15.7$ & $1.33$ \\
18--22 & $19.7$ & $0.30$ & $19.7$ & $0.89$ & $19.6$ & $1.33$ \\
22--30 & $25.0$ & $0.31$ & $25.1$ & $0.88$ & $25.1$ & $1.34$ \\
30--50 & $35.4$ & $0.31$ & $35.7$ & $0.89$ & $36.0$ & $1.34$ \\
\hline
\end{tabular}
\label{tab:MeanKinematics}
\end{table}

\clearpage

\begin{table}[p]
\centering
\caption{\lth\ in the CS frame for the \PsiOne.}
\label{tab:first}
\begin{tabular}{ccccccc}
\hline\noalign{\smallskip}
\multirow{2}{*}{$|y|$} & \pt & \multirow{2}{*}{\lth} & \multicolumn{3}{c}{Total uncertainty} & Stat.\ unc. \\
 & (\GeVns) & & 68.3\% CL & 95.5\% CL & 99.7\% CL & 68.3\% CL \\ \noalign{\smallskip}
\hline\noalign{\smallskip}

\multirow{12}{*}{0.0--0.6}
& 14--16 & $ -0.116 $ & \range{-0.116}{+0.102} & \range{-0.224}{+0.210} & \range{-0.333}{+0.319} & \range{-0.025}{+0.024} \\ \noalign{\smallskip}
& 16--18 & $ -0.062 $ & \range{-0.088}{+0.082} & \range{-0.173}{+0.167} & \range{-0.258}{+0.251} & \range{-0.024}{+0.024} \\ \noalign{\smallskip}
& 18--20 & $ -0.102 $ & \range{-0.073}{+0.066} & \range{-0.139}{+0.135} & \range{-0.208}{+0.201} & \range{-0.026}{+0.025} \\ \noalign{\smallskip}
& 20--22 & $ -0.066 $ & \range{-0.063}{+0.058} & \range{-0.126}{+0.121} & \range{-0.187}{+0.181} & \range{-0.028}{+0.028} \\ \noalign{\smallskip}
& 22--25 & $ -0.107 $ & \range{-0.061}{+0.055} & \range{-0.117}{+0.114} & \range{-0.175}{+0.169} & \range{-0.027}{+0.027} \\ \noalign{\smallskip}
& 25--30 & $ -0.049 $ & \range{-0.088}{+0.080} & \range{-0.172}{+0.163} & \range{-0.260}{+0.252} & \range{-0.029}{+0.029} \\
\end{tabular}
\end{table}

\begin{table}[p]
\centering
\caption{\lph\ in the CS frame for the \PsiOne.}
\begin{tabular}{ccccccc}
\hline\noalign{\smallskip}
\multirow{2}{*}{$|y|$} & \pt & \multirow{2}{*}{\lph} & \multicolumn{3}{c}{Total uncertainty} & Stat.\ unc. \\
 & (\GeVns) & & 68.3\% CL & 95.5\% CL & 99.7\% CL & 68.3\% CL \\ \noalign{\smallskip}
\hline\noalign{\smallskip}

\multirow{12}{*}{0.0--0.6}
& 14--16 & $ +0.087 $ & \range{-0.112}{+0.099} & \range{-0.217}{+0.204} & \range{-0.322}{+0.309} & \range{-0.027}{+0.022} \\ \noalign{\smallskip}
& 16--18 & $ +0.024 $ & \range{-0.084}{+0.075} & \range{-0.165}{+0.156} & \range{-0.247}{+0.238} & \range{-0.027}{+0.022} \\ \noalign{\smallskip}
& 18--20 & $ +0.079 $ & \range{-0.073}{+0.064} & \range{-0.143}{+0.135} & \range{-0.210}{+0.202} & \range{-0.027}{+0.022} \\ \noalign{\smallskip}
& 20--22 & $ +0.028 $ & \range{-0.066}{+0.059} & \range{-0.128}{+0.118} & \range{-0.187}{+0.181} & \range{-0.031}{+0.025} \\ \noalign{\smallskip}
& 22--25 & $ +0.068 $ & \range{-0.062}{+0.054} & \range{-0.119}{+0.113} & \range{-0.179}{+0.170} & \range{-0.027}{+0.024} \\ \noalign{\smallskip}
& 25--30 & $ +0.020 $ & \range{-0.056}{+0.049} & \range{-0.110}{+0.103} & \range{-0.163}{+0.157} & \range{-0.027}{+0.023} \\
\end{tabular}
\end{table}

\begin{table}[p]
\centering
\caption{\ltp\ in the CS frame for the \PsiOne.}
\begin{tabular}{ccccccc}
\hline\noalign{\smallskip}
\multirow{2}{*}{$|y|$} & \pt & \multirow{2}{*}{\ltp} & \multicolumn{3}{c}{Total uncertainty} & Stat.\ unc. \\
 & (\GeVns) & & 68.3\% CL & 95.5\% CL & 99.7\% CL & 68.3\% CL \\ \noalign{\smallskip}
\hline\noalign{\smallskip}

\multirow{12}{*}{0.0--0.6}
& 14--16 & $ +0.001 $ & \range{-0.034}{+0.031} & \range{-0.066}{+0.063} & \range{-0.099}{+0.094} & \range{-0.009}{+0.008} \\ \noalign{\smallskip}
& 16--18 & $ -0.030 $ & \range{-0.034}{+0.031} & \range{-0.069}{+0.065} & \range{-0.101}{+0.097} & \range{-0.010}{+0.009} \\ \noalign{\smallskip}
& 18--20 & $ -0.017 $ & \range{-0.035}{+0.031} & \range{-0.068}{+0.064} & \range{-0.100}{+0.096} & \range{-0.011}{+0.011} \\ \noalign{\smallskip}
& 20--22 & $ -0.001 $ & \range{-0.034}{+0.030} & \range{-0.066}{+0.062} & \range{-0.098}{+0.093} & \range{-0.013}{+0.012} \\ \noalign{\smallskip}
& 22--25 & $ -0.009 $ & \range{-0.030}{+0.026} & \range{-0.057}{+0.054} & \range{-0.085}{+0.081} & \range{-0.014}{+0.012} \\ \noalign{\smallskip}
& 25--30 & $ +0.003 $ & \range{-0.029}{+0.026} & \range{-0.056}{+0.054} & \range{-0.085}{+0.082} & \range{-0.015}{+0.014} \\
\end{tabular}
\end{table}

\begin{table}[p]
\centering
\caption{\ltilde\ in the CS frame for the \PsiOne.}
\begin{tabular}{ccccccc}
\hline\noalign{\smallskip}
\multirow{2}{*}{$|y|$} & \pt & \multirow{2}{*}{\ltilde} & \multicolumn{3}{c}{Total uncertainty} & Stat.\ unc. \\
 & (\GeVns) & & 68.3\% CL & 95.5\% CL & 99.7\% CL & 68.3\% CL \\ \noalign{\smallskip}
\hline\noalign{\smallskip}

\multirow{12}{*}{0.0--0.6}
& 14--16 & $ +0.159 $ & \range{-0.234}{+0.208} & \range{-0.454}{+0.428} & \range{-0.675}{+0.649} & \range{-0.063}{+0.057} \\ \noalign{\smallskip}
& 16--18 & $ +0.011 $ & \range{-0.168}{+0.145} & \range{-0.321}{+0.306} & \range{-0.481}{+0.458} & \range{-0.058}{+0.051} \\ \noalign{\smallskip}
& 18--20 & $ +0.147 $ & \range{-0.153}{+0.134} & \range{-0.297}{+0.285} & \range{-0.441}{+0.429} & \range{-0.062}{+0.059} \\ \noalign{\smallskip}
& 20--22 & $ +0.019 $ & \range{-0.144}{+0.135} & \range{-0.286}{+0.277} & \range{-0.429}{+0.413} & \range{-0.065}{+0.063} \\ \noalign{\smallskip}
& 22--25 & $ +0.104 $ & \range{-0.124}{+0.116} & \range{-0.244}{+0.236} & \range{-0.364}{+0.356} & \range{-0.066}{+0.066} \\ \noalign{\smallskip}
& 25--30 & $ +0.013 $ & \range{-0.105}{+0.097} & \range{-0.206}{+0.199} & \range{-0.307}{+0.300} & \range{-0.063}{+0.060} \\
\end{tabular}
\end{table}

\clearpage
\begin{table}[p]
\centering
\caption{\lth\ in the HX frame for the \PsiOne.}
\begin{tabular}{ccccccc}
\hline\noalign{\smallskip}
\multirow{2}{*}{$|y|$} & \pt & \multirow{2}{*}{\lth} & \multicolumn{3}{c}{Total uncertainty} & Stat.\ unc. \\
 & (\GeVns) & & 68.3\% CL & 95.5\% CL & 99.7\% CL & 68.3\% CL \\ \noalign{\smallskip}
\hline\noalign{\smallskip}

\multirow{12}{*}{0.0--0.6}
& 14--16 & $ +0.197 $ & \range{-0.228}{+0.208} & \range{-0.439}{+0.419} & \range{-0.651}{+0.630} & \range{-0.061}{+0.055} \\ \noalign{\smallskip}
& 16--18 & $ +0.070 $ & \range{-0.167}{+0.153} & \range{-0.336}{+0.313} & \range{-0.496}{+0.473} & \range{-0.055}{+0.051} \\ \noalign{\smallskip}
& 18--20 & $ +0.187 $ & \range{-0.143}{+0.129} & \range{-0.273}{+0.259} & \range{-0.409}{+0.395} & \range{-0.059}{+0.053} \\ \noalign{\smallskip}
& 20--22 & $ +0.078 $ & \range{-0.124}{+0.116} & \range{-0.247}{+0.233} & \range{-0.364}{+0.356} & \range{-0.059}{+0.055} \\ \noalign{\smallskip}
& 22--25 & $ +0.172 $ & \range{-0.110}{+0.103} & \range{-0.220}{+0.212} & \range{-0.329}{+0.316} & \range{-0.061}{+0.058} \\ \noalign{\smallskip}
& 25--30 & $ +0.056 $ & \range{-0.100}{+0.090} & \range{-0.198}{+0.188} & \range{-0.295}{+0.286} & \range{-0.054}{+0.050} \\
\end{tabular}
\end{table}

\begin{table}[p]
\centering
\caption{\lph\ in the HX frame for the \PsiOne.}
\begin{tabular}{ccccccc}
\hline\noalign{\smallskip}
\multirow{2}{*}{$|y|$} & \pt & \multirow{2}{*}{\lph} & \multicolumn{3}{c}{Total uncertainty} & Stat.\ unc. \\
 & (\GeVns) & & 68.3\% CL & 95.5\% CL & 99.7\% CL & 68.3\% CL \\ \noalign{\smallskip}
\hline\noalign{\smallskip}

\multirow{12}{*}{0.0--0.6}
& 14--16 & $ -0.016 $ & \range{-0.027}{+0.025} & \range{-0.053}{+0.051} & \range{-0.079}{+0.077} & \range{-0.005}{+0.004} \\ \noalign{\smallskip}
& 16--18 & $ -0.021 $ & \range{-0.029}{+0.026} & \range{-0.056}{+0.054} & \range{-0.084}{+0.082} & \range{-0.006}{+0.005} \\ \noalign{\smallskip}
& 18--20 & $ -0.014 $ & \range{-0.025}{+0.023} & \range{-0.050}{+0.047} & \range{-0.074}{+0.072} & \range{-0.007}{+0.006} \\ \noalign{\smallskip}
& 20--22 & $ -0.020 $ & \range{-0.032}{+0.030} & \range{-0.063}{+0.059} & \range{-0.093}{+0.090} & \range{-0.009}{+0.008} \\ \noalign{\smallskip}
& 22--25 & $ -0.022 $ & \range{-0.031}{+0.027} & \range{-0.058}{+0.056} & \range{-0.088}{+0.084} & \range{-0.010}{+0.009} \\ \noalign{\smallskip}
& 25--30 & $ -0.015 $ & \range{-0.042}{+0.039} & \range{-0.081}{+0.077} & \range{-0.121}{+0.118} & \range{-0.011}{+0.010} \\
\end{tabular}
\end{table}

\begin{table}[p]
\centering
\caption{\ltp\ in the HX frame for the \PsiOne.}
\begin{tabular}{ccccccc}
\hline\noalign{\smallskip}
\multirow{2}{*}{$|y|$} & \pt & \multirow{2}{*}{\ltp} & \multicolumn{3}{c}{Total uncertainty} & Stat.\ unc. \\
 & (\GeVns) & & 68.3\% CL & 95.5\% CL & 99.7\% CL & 68.3\% CL \\ \noalign{\smallskip}
\hline\noalign{\smallskip}

\multirow{12}{*}{0.0--0.6}
& 14--16 & $ -0.012 $ & \range{-0.029}{+0.025} & \range{-0.055}{+0.053} & \range{-0.082}{+0.079} & \range{-0.011}{+0.010} \\ \noalign{\smallskip}
& 16--18 & $ +0.027 $ & \range{-0.032}{+0.029} & \range{-0.063}{+0.060} & \range{-0.094}{+0.091} & \range{-0.011}{+0.010} \\ \noalign{\smallskip}
& 18--20 & $ +0.011 $ & \range{-0.032}{+0.029} & \range{-0.062}{+0.057} & \range{-0.092}{+0.088} & \range{-0.013}{+0.011} \\ \noalign{\smallskip}
& 20--22 & $ -0.003 $ & \range{-0.032}{+0.030} & \range{-0.063}{+0.059} & \range{-0.094}{+0.090} & \range{-0.014}{+0.013} \\ \noalign{\smallskip}
& 22--25 & $ +0.004 $ & \range{-0.029}{+0.026} & \range{-0.057}{+0.054} & \range{-0.084}{+0.080} & \range{-0.015}{+0.014} \\ \noalign{\smallskip}
& 25--30 & $ -0.005 $ & \range{-0.032}{+0.029} & \range{-0.062}{+0.059} & \range{-0.093}{+0.090} & \range{-0.016}{+0.014} \\
\end{tabular}
\end{table}

\begin{table}[p]
\centering
\caption{\ltilde\ in the HX frame for the \PsiOne.}
\begin{tabular}{ccccccc}
\hline\noalign{\smallskip}
\multirow{2}{*}{$|y|$} & \pt & \multirow{2}{*}{\ltilde} & \multicolumn{3}{c}{Total uncertainty} & Stat.\ unc. \\
 & (\GeVns) & & 68.3\% CL & 95.5\% CL & 99.7\% CL & 68.3\% CL \\ \noalign{\smallskip}
\hline\noalign{\smallskip}

\multirow{12}{*}{0.0--0.6}
& 14--16 & $ +0.147 $ & \range{-0.236}{+0.209} & \range{-0.464}{+0.437} & \range{-0.692}{+0.665} & \range{-0.064}{+0.056} \\ \noalign{\smallskip}
& 16--18 & $ +0.006 $ & \range{-0.178}{+0.152} & \range{-0.343}{+0.326} & \range{-0.509}{+0.491} & \range{-0.057}{+0.051} \\ \noalign{\smallskip}
& 18--20 & $ +0.144 $ & \range{-0.151}{+0.139} & \range{-0.299}{+0.280} & \range{-0.440}{+0.428} & \range{-0.060}{+0.055} \\ \noalign{\smallskip}
& 20--22 & $ +0.018 $ & \range{-0.136}{+0.124} & \range{-0.266}{+0.254} & \range{-0.397}{+0.377} & \range{-0.064}{+0.060} \\ \noalign{\smallskip}
& 22--25 & $ +0.105 $ & \range{-0.127}{+0.116} & \range{-0.245}{+0.240} & \range{-0.369}{+0.358} & \range{-0.066}{+0.063} \\ \noalign{\smallskip}
& 25--30 & $ +0.013 $ & \range{-0.106}{+0.098} & \range{-0.211}{+0.203} & \range{-0.310}{+0.302} & \range{-0.060}{+0.058} \\
\end{tabular}
\end{table}

\clearpage
\begin{table}[p]
\centering
\caption{\lth\ in the PX frame for the \PsiOne.}
\begin{tabular}{ccccccc}
\hline\noalign{\smallskip}
\multirow{2}{*}{$|y|$} & \pt & \multirow{2}{*}{\lth} & \multicolumn{3}{c}{Total uncertainty} & Stat.\ unc. \\
 & (\GeVns) & & 68.3\% CL & 95.5\% CL & 99.7\% CL & 68.3\% CL \\ \noalign{\smallskip}
\hline\noalign{\smallskip}

\multirow{12}{*}{0.0--0.6}
& 14--16 & $ +0.215 $ & \range{-0.227}{+0.208} & \range{-0.444}{+0.425} & \range{-0.661}{+0.642} & \range{-0.062}{+0.058} \\ \noalign{\smallskip}
& 16--18 & $ +0.076 $ & \range{-0.173}{+0.155} & \range{-0.337}{+0.319} & \range{-0.501}{+0.475} & \range{-0.055}{+0.049} \\ \noalign{\smallskip}
& 18--20 & $ +0.193 $ & \range{-0.148}{+0.134} & \range{-0.289}{+0.275} & \range{-0.438}{+0.423} & \range{-0.059}{+0.056} \\ \noalign{\smallskip}
& 20--22 & $ +0.086 $ & \range{-0.128}{+0.114} & \range{-0.246}{+0.233} & \range{-0.365}{+0.351} & \range{-0.062}{+0.057} \\ \noalign{\smallskip}
& 22--25 & $ +0.177 $ & \range{-0.119}{+0.108} & \range{-0.232}{+0.221} & \range{-0.346}{+0.335} & \range{-0.061}{+0.057} \\ \noalign{\smallskip}
& 25--30 & $ +0.062 $ & \range{-0.101}{+0.093} & \range{-0.201}{+0.193} & \range{-0.301}{+0.293} & \range{-0.052}{+0.051} \\
\end{tabular}
\end{table}

\begin{table}[p]
\centering
\caption{\lph\ in the PX frame for the \PsiOne.}
\begin{tabular}{ccccccc}
\hline\noalign{\smallskip}
\multirow{2}{*}{$|y|$} & \pt & \multirow{2}{*}{\lph} & \multicolumn{3}{c}{Total uncertainty} & Stat.\ unc. \\
 & (\GeVns) & & 68.3\% CL & 95.5\% CL & 99.7\% CL & 68.3\% CL \\ \noalign{\smallskip}
\hline\noalign{\smallskip}

\multirow{12}{*}{0.0--0.6}
& 14--16 & $ -0.016 $ & \range{-0.025}{+0.022} & \range{-0.048}{+0.047} & \range{-0.073}{+0.070} & \range{-0.005}{+0.004} \\ \noalign{\smallskip}
& 16--18 & $ -0.020 $ & \range{-0.028}{+0.025} & \range{-0.055}{+0.052} & \range{-0.082}{+0.078} & \range{-0.006}{+0.006} \\ \noalign{\smallskip}
& 18--20 & $ -0.013 $ & \range{-0.026}{+0.023} & \range{-0.049}{+0.048} & \range{-0.074}{+0.072} & \range{-0.007}{+0.007} \\ \noalign{\smallskip}
& 20--22 & $ -0.020 $ & \range{-0.031}{+0.027} & \range{-0.059}{+0.056} & \range{-0.087}{+0.084} & \range{-0.009}{+0.008} \\ \noalign{\smallskip}
& 22--25 & $ -0.022 $ & \range{-0.029}{+0.026} & \range{-0.055}{+0.053} & \range{-0.083}{+0.079} & \range{-0.010}{+0.009} \\ \noalign{\smallskip}
& 25--30 & $ -0.015 $ & \range{-0.042}{+0.037} & \range{-0.082}{+0.077} & \range{-0.120}{+0.118} & \range{-0.011}{+0.010} \\
\end{tabular}
\end{table}

\begin{table}[p]
\centering
\caption{\ltp\ in the PX frame for the \PsiOne.}
\begin{tabular}{ccccccc}
\hline\noalign{\smallskip}
\multirow{2}{*}{$|y|$} & \pt & \multirow{2}{*}{\ltp} & \multicolumn{3}{c}{Total uncertainty} & Stat.\ unc. \\
 & (\GeVns) & & 68.3\% CL & 95.5\% CL & 99.7\% CL & 68.3\% CL \\ \noalign{\smallskip}
\hline\noalign{\smallskip}

\multirow{12}{*}{0.0--0.6}
& 14--16 & $ +0.000 $ & \range{-0.035}{+0.032} & \range{-0.068}{+0.066} & \range{-0.102}{+0.099} & \range{-0.010}{+0.010} \\ \noalign{\smallskip}
& 16--18 & $ +0.031 $ & \range{-0.034}{+0.031} & \range{-0.067}{+0.064} & \range{-0.099}{+0.096} & \range{-0.011}{+0.010} \\ \noalign{\smallskip}
& 18--20 & $ +0.019 $ & \range{-0.035}{+0.032} & \range{-0.067}{+0.063} & \range{-0.100}{+0.097} & \range{-0.012}{+0.011} \\ \noalign{\smallskip}
& 20--22 & $ +0.001 $ & \range{-0.033}{+0.030} & \range{-0.066}{+0.061} & \range{-0.097}{+0.094} & \range{-0.014}{+0.013} \\ \noalign{\smallskip}
& 22--25 & $ +0.010 $ & \range{-0.030}{+0.027} & \range{-0.057}{+0.054} & \range{-0.084}{+0.083} & \range{-0.014}{+0.013} \\ \noalign{\smallskip}
& 25--30 & $ -0.003 $ & \range{-0.030}{+0.028} & \range{-0.059}{+0.057} & \range{-0.088}{+0.086} & \range{-0.016}{+0.014} \\
\end{tabular}
\end{table}

\begin{table}[p]
\centering
\caption{\ltilde\ in the PX frame for the \PsiOne.}
\begin{tabular}{ccccccc}
\hline\noalign{\smallskip}
\multirow{2}{*}{$|y|$} & \pt & \multirow{2}{*}{\ltilde} & \multicolumn{3}{c}{Total uncertainty} & Stat.\ unc. \\
 & (\GeVns) & & 68.3\% CL & 95.5\% CL & 99.7\% CL & 68.3\% CL \\ \noalign{\smallskip}
\hline\noalign{\smallskip}

\multirow{12}{*}{0.0--0.6}
& 14--16 & $ +0.163 $ & \range{-0.238}{+0.212} & \range{-0.451}{+0.425} & \range{-0.676}{+0.650} & \range{-0.062}{+0.059} \\ \noalign{\smallskip}
& 16--18 & $ +0.016 $ & \range{-0.163}{+0.148} & \range{-0.327}{+0.312} & \range{-0.483}{+0.467} & \range{-0.058}{+0.051} \\ \noalign{\smallskip}
& 18--20 & $ +0.153 $ & \range{-0.148}{+0.136} & \range{-0.297}{+0.278} & \range{-0.439}{+0.420} & \range{-0.060}{+0.056} \\ \noalign{\smallskip}
& 20--22 & $ +0.025 $ & \range{-0.146}{+0.131} & \range{-0.281}{+0.266} & \range{-0.416}{+0.400} & \range{-0.067}{+0.061} \\ \noalign{\smallskip}
& 22--25 & $ +0.110 $ & \range{-0.134}{+0.120} & \range{-0.261}{+0.247} & \range{-0.388}{+0.374} & \range{-0.066}{+0.062} \\ \noalign{\smallskip}
& 25--30 & $ +0.017 $ & \range{-0.108}{+0.094} & \range{-0.204}{+0.196} & \range{-0.306}{+0.297} & \range{-0.062}{+0.060} \\
\end{tabular}
\end{table}

\clearpage
\begin{table}[p]
\centering
\caption{\lth\ in the CS frame for the \PsiTwo.}
\begin{tabular}{ccccccc}
\hline\noalign{\smallskip}
\multirow{2}{*}{$|y|$} & \pt & \multirow{2}{*}{\lth} & \multicolumn{3}{c}{Total uncertainty} & Stat.\ unc. \\
 & (\GeVns) & & 68.3\% CL & 95.5\% CL & 99.7\% CL & 68.3\% CL \\ \noalign{\smallskip}
\hline\noalign{\smallskip}

\multirow{5}{*}{0.0--0.6}
& 14--18 & $ -0.124 $ & \range{-0.108}{+0.105} & \range{-0.215}{+0.224} & \range{-0.315}{+0.342} & \range{-0.063}{+0.078} \\ \noalign{\smallskip}
& 18--22 & $ -0.118 $ & \range{-0.104}{+0.107} & \range{-0.196}{+0.215} & \range{-0.288}{+0.339} & \range{-0.076}{+0.086} \\ \noalign{\smallskip}
& 22--30 & $ -0.009 $ & \range{-0.105}{+0.113} & \range{-0.199}{+0.237} & \range{-0.294}{+0.367} & \range{-0.087}{+0.103} \\ \noalign{\smallskip}
& 30--50 & $ +0.014 $ & \range{-0.133}{+0.163} & \range{-0.256}{+0.346} & \range{-0.378}{+0.573} & \range{-0.116}{+0.154} \\ \noalign{\smallskip}
\hline\noalign{\smallskip}
\multirow{5}{*}{0.6--1.2}
& 14--18 & $ -0.002 $ & \range{-0.091}{+0.092} & \range{-0.177}{+0.194} & \range{-0.263}{+0.295} & \range{-0.064}{+0.075} \\ \noalign{\smallskip}
& 18--22 & $ +0.048 $ & \range{-0.101}{+0.118} & \range{-0.201}{+0.242} & \range{-0.290}{+0.390} & \range{-0.081}{+0.103} \\
\end{tabular}
\end{table}

\begin{table}[p]
\centering
\caption{\lph\ in the CS frame for the \PsiTwo.}
\begin{tabular}{ccccccc}
\hline\noalign{\smallskip}
\multirow{2}{*}{$|y|$} & \pt & \multirow{2}{*}{\lph} & \multicolumn{3}{c}{Total uncertainty} & Stat.\ unc. \\
 & (\GeVns) & & 68.3\% CL & 95.5\% CL & 99.7\% CL & 68.3\% CL \\ \noalign{\smallskip}
\hline\noalign{\smallskip}

\multirow{5}{*}{0.0--0.6}
& 14--18 & $ +0.084 $ & \range{-0.124}{+0.098} & \range{-0.241}{+0.210} & \range{-0.363}{+0.309} & \range{-0.087}{+0.057} \\ \noalign{\smallskip}
& 18--22 & $ +0.075 $ & \range{-0.109}{+0.084} & \range{-0.220}{+0.172} & \range{-0.336}{+0.255} & \range{-0.094}{+0.064} \\ \noalign{\smallskip}
& 22--30 & $ -0.038 $ & \range{-0.111}{+0.087} & \range{-0.217}{+0.177} & \range{-0.334}{+0.263} & \range{-0.099}{+0.071} \\ \noalign{\smallskip}
& 30--50 & $ +0.018 $ & \range{-0.137}{+0.095} & \range{-0.274}{+0.197} & \range{-0.431}{+0.293} & \range{-0.127}{+0.083} \\ \noalign{\smallskip}
\hline\noalign{\smallskip}
\multirow{5}{*}{0.6--1.2}
& 14--18 & $ -0.076 $ & \range{-0.103}{+0.088} & \range{-0.209}{+0.179} & \range{-0.315}{+0.264} & \range{-0.083}{+0.058} \\ \noalign{\smallskip}
& 18--22 & $ -0.155 $ & \range{-0.125}{+0.096} & \range{-0.244}{+0.197} & \range{-0.376}{+0.287} & \range{-0.106}{+0.075} \\
\end{tabular}
\end{table}

\begin{table}[p]
\centering
\caption{\ltp\ in the CS frame for the \PsiTwo.}
\begin{tabular}{ccccccc}
\hline\noalign{\smallskip}
\multirow{2}{*}{$|y|$} & \pt & \multirow{2}{*}{\ltp} & \multicolumn{3}{c}{Total uncertainty} & Stat.\ unc. \\
 & (\GeVns) & & 68.3\% CL & 95.5\% CL & 99.7\% CL & 68.3\% CL \\ \noalign{\smallskip}
\hline\noalign{\smallskip}

\multirow{5}{*}{0.0--0.6}
& 14--18 & $ +0.012 $ & \range{-0.045}{+0.042} & \range{-0.089}{+0.088} & \range{-0.131}{+0.131} & \range{-0.029}{+0.029} \\ \noalign{\smallskip}
& 18--22 & $ -0.098 $ & \range{-0.046}{+0.040} & \range{-0.090}{+0.082} & \range{-0.135}{+0.124} & \range{-0.039}{+0.033} \\ \noalign{\smallskip}
& 22--30 & $ -0.029 $ & \range{-0.053}{+0.049} & \range{-0.104}{+0.097} & \range{-0.152}{+0.148} & \range{-0.046}{+0.042} \\ \noalign{\smallskip}
& 30--50 & $ +0.006 $ & \range{-0.078}{+0.069} & \range{-0.149}{+0.144} & \range{-0.220}{+0.219} & \range{-0.070}{+0.067} \\ \noalign{\smallskip}
\hline\noalign{\smallskip}
\multirow{5}{*}{0.6--1.2}
& 14--18 & $ -0.038 $ & \range{-0.048}{+0.045} & \range{-0.098}{+0.092} & \range{-0.144}{+0.141} & \range{-0.029}{+0.027} \\ \noalign{\smallskip}
& 18--22 & $ -0.025 $ & \range{-0.052}{+0.046} & \range{-0.099}{+0.096} & \range{-0.149}{+0.144} & \range{-0.040}{+0.037} \\
\end{tabular}
\end{table}

\begin{table}[p]
\centering
\caption{\ltilde\ in the CS frame for the \PsiTwo.}
\begin{tabular}{ccccccc}
\hline\noalign{\smallskip}
\multirow{2}{*}{$|y|$} & \pt & \multirow{2}{*}{\ltilde} & \multicolumn{3}{c}{Total uncertainty} & Stat.\ unc. \\
 & (\GeVns) & & 68.3\% CL & 95.5\% CL & 99.7\% CL & 68.3\% CL \\ \noalign{\smallskip}
\hline\noalign{\smallskip}

\multirow{5}{*}{0.0--0.6}
& 14--18 & $ +0.139 $ & \range{-0.257}{+0.232} & \range{-0.495}{+0.482} & \range{-0.732}{+0.733} & \range{-0.178}{+0.172} \\ \noalign{\smallskip}
& 18--22 & $ +0.108 $ & \range{-0.230}{+0.225} & \range{-0.453}{+0.458} & \range{-0.664}{+0.692} & \range{-0.191}{+0.191} \\ \noalign{\smallskip}
& 22--30 & $ -0.129 $ & \range{-0.193}{+0.207} & \range{-0.387}{+0.423} & \range{-0.560}{+0.650} & \range{-0.171}{+0.190} \\ \noalign{\smallskip}
& 30--50 & $ +0.020 $ & \range{-0.250}{+0.321} & \range{-0.500}{+0.678} & \range{-0.714}{+1.160} & \range{-0.238}{+0.316} \\ \noalign{\smallskip}
\hline\noalign{\smallskip}
\multirow{5}{*}{0.6--1.2}
& 14--18 & $ -0.214 $ & \range{-0.186}{+0.176} & \range{-0.367}{+0.367} & \range{-0.548}{+0.557} & \range{-0.128}{+0.126} \\ \noalign{\smallskip}
& 18--22 & $ -0.365 $ & \range{-0.183}{+0.179} & \range{-0.355}{+0.361} & \range{-0.509}{+0.551} & \range{-0.141}{+0.149} \\
\end{tabular}
\end{table}

\clearpage
\begin{table}[p]
\centering
\caption{\lth\ in the HX frame for the \PsiTwo.}
\begin{tabular}{ccccccc}
\hline\noalign{\smallskip}
\multirow{2}{*}{$|y|$} & \pt & \multirow{2}{*}{\lth} & \multicolumn{3}{c}{Total uncertainty} & Stat.\ unc. \\
 & (\GeVns) & & 68.3\% CL & 95.5\% CL & 99.7\% CL & 68.3\% CL \\ \noalign{\smallskip}
\hline\noalign{\smallskip}

\multirow{5}{*}{0.0--0.6}
& 14--18 & $ +0.260 $ & \range{-0.243}{+0.232} & \range{-0.481}{+0.470} & \range{-0.719}{+0.707} & \range{-0.173}{+0.165} \\ \noalign{\smallskip}
& 18--22 & $ +0.225 $ & \range{-0.223}{+0.204} & \range{-0.425}{+0.429} & \range{-0.616}{+0.643} & \range{-0.183}{+0.180} \\ \noalign{\smallskip}
& 22--30 & $ -0.026 $ & \range{-0.193}{+0.182} & \range{-0.366}{+0.384} & \range{-0.538}{+0.585} & \range{-0.155}{+0.162} \\ \noalign{\smallskip}
& 30--50 & $ +0.074 $ & \range{-0.230}{+0.220} & \range{-0.444}{+0.457} & \range{-0.657}{+0.718} & \range{-0.209}{+0.208} \\ \noalign{\smallskip}
\hline\noalign{\smallskip}
\multirow{5}{*}{0.6--1.2}
& 14--18 & $ -0.064 $ & \range{-0.164}{+0.154} & \range{-0.324}{+0.322} & \range{-0.475}{+0.481} & \range{-0.133}{+0.127} \\ \noalign{\smallskip}
& 18--22 & $ -0.188 $ & \range{-0.169}{+0.165} & \range{-0.336}{+0.332} & \range{-0.485}{+0.508} & \range{-0.147}{+0.142} \\
\end{tabular}
\end{table}

\begin{table}[p]
\centering
\caption{\lph\ in the HX frame for the \PsiTwo.}
\begin{tabular}{ccccccc}
\hline\noalign{\smallskip}
\multirow{2}{*}{$|y|$} & \pt & \multirow{2}{*}{\lph} & \multicolumn{3}{c}{Total uncertainty} & Stat.\ unc. \\
 & (\GeVns) & & 68.3\% CL & 95.5\% CL & 99.7\% CL & 68.3\% CL \\ \noalign{\smallskip}
\hline\noalign{\smallskip}

\multirow{5}{*}{0.0--0.6}
& 14--18 & $ -0.021 $ & \range{-0.033}{+0.030} & \range{-0.065}{+0.061} & \range{-0.098}{+0.094} & \range{-0.015}{+0.013} \\ \noalign{\smallskip}
& 18--22 & $ -0.030 $ & \range{-0.039}{+0.035} & \range{-0.077}{+0.073} & \range{-0.115}{+0.111} & \range{-0.025}{+0.024} \\ \noalign{\smallskip}
& 22--30 & $ -0.024 $ & \range{-0.044}{+0.039} & \range{-0.086}{+0.081} & \range{-0.126}{+0.122} & \range{-0.033}{+0.030} \\ \noalign{\smallskip}
& 30--50 & $ +0.010 $ & \range{-0.069}{+0.065} & \range{-0.131}{+0.130} & \range{-0.193}{+0.203} & \range{-0.060}{+0.057} \\ \noalign{\smallskip}
\hline\noalign{\smallskip}
\multirow{5}{*}{0.6--1.2}
& 14--18 & $ -0.045 $ & \range{-0.039}{+0.035} & \range{-0.078}{+0.074} & \range{-0.115}{+0.111} & \range{-0.015}{+0.014} \\ \noalign{\smallskip}
& 18--22 & $ -0.052 $ & \range{-0.040}{+0.036} & \range{-0.079}{+0.074} & \range{-0.117}{+0.115} & \range{-0.023}{+0.022} \\
\end{tabular}
\end{table}

\begin{table}[p]
\centering
\caption{\ltp\ in the HX frame for the \PsiTwo.}
\begin{tabular}{ccccccc}
\hline\noalign{\smallskip}
\multirow{2}{*}{$|y|$} & \pt & \multirow{2}{*}{\ltp} & \multicolumn{3}{c}{Total uncertainty} & Stat.\ unc. \\
 & (\GeVns) & & 68.3\% CL & 95.5\% CL & 99.7\% CL & 68.3\% CL \\ \noalign{\smallskip}
\hline\noalign{\smallskip}

\multirow{5}{*}{0.0--0.6}
& 14--18 & $ -0.030 $ & \range{-0.054}{+0.046} & \range{-0.105}{+0.095} & \range{-0.154}{+0.144} & \range{-0.036}{+0.029} \\ \noalign{\smallskip}
& 18--22 & $ +0.096 $ & \range{-0.048}{+0.043} & \range{-0.093}{+0.089} & \range{-0.138}{+0.134} & \range{-0.040}{+0.037} \\ \noalign{\smallskip}
& 22--30 & $ +0.030 $ & \range{-0.053}{+0.047} & \range{-0.102}{+0.096} & \range{-0.151}{+0.145} & \range{-0.044}{+0.041} \\ \noalign{\smallskip}
& 30--50 & $ -0.007 $ & \range{-0.078}{+0.067} & \range{-0.156}{+0.141} & \range{-0.234}{+0.212} & \range{-0.075}{+0.064} \\ \noalign{\smallskip}
\hline\noalign{\smallskip}
\multirow{5}{*}{0.6--1.2}
& 14--18 & $ +0.046 $ & \range{-0.053}{+0.047} & \range{-0.102}{+0.096} & \range{-0.151}{+0.147} & \range{-0.029}{+0.028} \\ \noalign{\smallskip}
& 18--22 & $ +0.044 $ & \range{-0.053}{+0.048} & \range{-0.103}{+0.097} & \range{-0.152}{+0.147} & \range{-0.037}{+0.034} \\
\end{tabular}
\end{table}

\begin{table}[p]
\centering
\caption{\ltilde\ in the HX frame for the \PsiTwo.}
\begin{tabular}{ccccccc}
\hline\noalign{\smallskip}
\multirow{2}{*}{$|y|$} & \pt & \multirow{2}{*}{\ltilde} & \multicolumn{3}{c}{Total uncertainty} & Stat.\ unc. \\
 & (\GeVns) & & 68.3\% CL & 95.5\% CL & 99.7\% CL & 68.3\% CL \\ \noalign{\smallskip}
\hline\noalign{\smallskip}

\multirow{5}{*}{0.0--0.6}
& 14--18 & $ +0.193 $ & \range{-0.252}{+0.226} & \range{-0.485}{+0.471} & \range{-0.718}{+0.717} & \range{-0.171}{+0.171} \\ \noalign{\smallskip}
& 18--22 & $ +0.133 $ & \range{-0.226}{+0.215} & \range{-0.447}{+0.448} & \range{-0.657}{+0.681} & \range{-0.188}{+0.182} \\ \noalign{\smallskip}
& 22--30 & $ -0.096 $ & \range{-0.203}{+0.205} & \range{-0.385}{+0.419} & \range{-0.556}{+0.633} & \range{-0.178}{+0.189} \\ \noalign{\smallskip}
& 30--50 & $ +0.087 $ & \range{-0.273}{+0.329} & \range{-0.530}{+0.690} & \range{-0.754}{+1.154} & \range{-0.256}{+0.320} \\ \noalign{\smallskip}
\hline\noalign{\smallskip}
\multirow{5}{*}{0.6--1.2}
& 14--18 & $ -0.190 $ & \range{-0.176}{+0.171} & \range{-0.345}{+0.341} & \range{-0.515}{+0.519} & \range{-0.129}{+0.132} \\ \noalign{\smallskip}
& 18--22 & $ -0.329 $ & \range{-0.184}{+0.177} & \range{-0.351}{+0.353} & \range{-0.508}{+0.538} & \range{-0.145}{+0.147} \\
\end{tabular}
\end{table}

\clearpage
\begin{table}[p]
\centering
\caption{\lth\ in the PX frame for the \PsiTwo.}
\begin{tabular}{ccccccc}
\hline\noalign{\smallskip}
\multirow{2}{*}{$|y|$} & \pt & \multirow{2}{*}{\lth} & \multicolumn{3}{c}{Total uncertainty} & Stat.\ unc. \\
 & (\GeVns) & & 68.3\% CL & 95.5\% CL & 99.7\% CL & 68.3\% CL \\ \noalign{\smallskip}
\hline\noalign{\smallskip}

\multirow{5}{*}{0.0--0.6}
& 14--18 & $ +0.256 $ & \range{-0.244}{+0.225} & \range{-0.479}{+0.459} & \range{-0.701}{+0.694} & \range{-0.181}{+0.171} \\ \noalign{\smallskip}
& 18--22 & $ +0.234 $ & \range{-0.228}{+0.214} & \range{-0.431}{+0.429} & \range{-0.634}{+0.656} & \range{-0.185}{+0.181} \\ \noalign{\smallskip}
& 22--30 & $ -0.017 $ & \range{-0.187}{+0.185} & \range{-0.355}{+0.372} & \range{-0.523}{+0.567} & \range{-0.157}{+0.164} \\ \noalign{\smallskip}
& 30--50 & $ +0.070 $ & \range{-0.220}{+0.222} & \range{-0.436}{+0.460} & \range{-0.651}{+0.710} & \range{-0.212}{+0.209} \\ \noalign{\smallskip}
\hline\noalign{\smallskip}
\multirow{5}{*}{0.6--1.2}
& 14--18 & $ -0.078 $ & \range{-0.169}{+0.158} & \range{-0.328}{+0.317} & \range{-0.487}{+0.485} & \range{-0.130}{+0.120} \\ \noalign{\smallskip}
& 18--22 & $ -0.200 $ & \range{-0.175}{+0.165} & \range{-0.327}{+0.334} & \range{-0.478}{+0.504} & \range{-0.145}{+0.142} \\
\end{tabular}
\end{table}

\begin{table}[p]
\centering
\caption{\lph\ in the PX frame for the \PsiTwo.}
\begin{tabular}{ccccccc}
\hline\noalign{\smallskip}
\multirow{2}{*}{$|y|$} & \pt & \multirow{2}{*}{\lph} & \multicolumn{3}{c}{Total uncertainty} & Stat.\ unc. \\
 & (\GeVns) & & 68.3\% CL & 95.5\% CL & 99.7\% CL & 68.3\% CL \\ \noalign{\smallskip}
\hline\noalign{\smallskip}

\multirow{5}{*}{0.0--0.6}
& 14--18 & $ -0.022 $ & \range{-0.035}{+0.031} & \range{-0.067}{+0.063} & \range{-0.101}{+0.096} & \range{-0.015}{+0.013} \\ \noalign{\smallskip}
& 18--22 & $ -0.025 $ & \range{-0.039}{+0.035} & \range{-0.076}{+0.073} & \range{-0.112}{+0.110} & \range{-0.025}{+0.024} \\ \noalign{\smallskip}
& 22--30 & $ -0.023 $ & \range{-0.042}{+0.037} & \range{-0.082}{+0.077} & \range{-0.122}{+0.117} & \range{-0.033}{+0.030} \\ \noalign{\smallskip}
& 30--50 & $ +0.011 $ & \range{-0.068}{+0.064} & \range{-0.129}{+0.131} & \range{-0.190}{+0.202} & \range{-0.061}{+0.058} \\ \noalign{\smallskip}
\hline\noalign{\smallskip}
\multirow{5}{*}{0.6--1.2}
& 14--18 & $ -0.038 $ & \range{-0.036}{+0.033} & \range{-0.072}{+0.068} & \range{-0.107}{+0.103} & \range{-0.014}{+0.014} \\ \noalign{\smallskip}
& 18--22 & $ -0.048 $ & \range{-0.037}{+0.033} & \range{-0.074}{+0.070} & \range{-0.109}{+0.105} & \range{-0.023}{+0.022} \\
\end{tabular}
\end{table}

\begin{table}[p]
\centering
\caption{\ltp\ in the PX frame for the \PsiTwo.}
\begin{tabular}{ccccccc}
\hline\noalign{\smallskip}
\multirow{2}{*}{$|y|$} & \pt & \multirow{2}{*}{\ltp} & \multicolumn{3}{c}{Total uncertainty} & Stat.\ unc. \\
 & (\GeVns) & & 68.3\% CL & 95.5\% CL & 99.7\% CL & 68.3\% CL \\ \noalign{\smallskip}
\hline\noalign{\smallskip}

\multirow{5}{*}{0.0--0.6}
& 14--18 & $ -0.013 $ & \range{-0.049}{+0.043} & \range{-0.095}{+0.086} & \range{-0.141}{+0.132} & \range{-0.035}{+0.028} \\ \noalign{\smallskip}
& 18--22 & $ +0.108 $ & \range{-0.049}{+0.045} & \range{-0.094}{+0.090} & \range{-0.139}{+0.138} & \range{-0.040}{+0.036} \\ \noalign{\smallskip}
& 22--30 & $ +0.029 $ & \range{-0.054}{+0.048} & \range{-0.102}{+0.099} & \range{-0.150}{+0.149} & \range{-0.045}{+0.041} \\ \noalign{\smallskip}
& 30--50 & $ -0.007 $ & \range{-0.080}{+0.071} & \range{-0.154}{+0.142} & \range{-0.228}{+0.216} & \range{-0.075}{+0.063} \\ \noalign{\smallskip}
\hline\noalign{\smallskip}
\multirow{5}{*}{0.6--1.2}
& 14--18 & $ +0.039 $ & \range{-0.048}{+0.045} & \range{-0.096}{+0.090} & \range{-0.141}{+0.138} & \range{-0.029}{+0.026} \\ \noalign{\smallskip}
& 18--22 & $ +0.024 $ & \range{-0.048}{+0.044} & \range{-0.093}{+0.088} & \range{-0.136}{+0.136} & \range{-0.036}{+0.034} \\
\end{tabular}
\end{table}

\begin{table}[p]
\centering
\caption{\ltilde\ in the PX frame for the \PsiTwo.}
\label{tab:last}
\begin{tabular}{ccccccc}
\hline\noalign{\smallskip}
\multirow{2}{*}{$|y|$} & \pt & \multirow{2}{*}{\ltilde} & \multicolumn{3}{c}{Total uncertainty} & Stat.\ unc. \\
 & (\GeVns) & & 68.3\% CL & 95.5\% CL & 99.7\% CL & 68.3\% CL \\ \noalign{\smallskip}
\hline\noalign{\smallskip}

\multirow{5}{*}{0.0--0.6}
& 14--18 & $ +0.185 $ & \range{-0.258}{+0.228} & \range{-0.495}{+0.478} & \range{-0.732}{+0.728} & \range{-0.174}{+0.174} \\ \noalign{\smallskip}
& 18--22 & $ +0.158 $ & \range{-0.234}{+0.215} & \range{-0.452}{+0.445} & \range{-0.660}{+0.675} & \range{-0.192}{+0.193} \\ \noalign{\smallskip}
& 22--30 & $ -0.085 $ & \range{-0.198}{+0.205} & \range{-0.378}{+0.417} & \range{-0.559}{+0.629} & \range{-0.170}{+0.190} \\ \noalign{\smallskip}
& 30--50 & $ +0.084 $ & \range{-0.264}{+0.320} & \range{-0.522}{+0.681} & \range{-0.745}{+1.162} & \range{-0.245}{+0.324} \\ \noalign{\smallskip}
\hline\noalign{\smallskip}
\multirow{5}{*}{0.6--1.2}
& 14--18 & $ -0.185 $ & \range{-0.173}{+0.164} & \range{-0.347}{+0.338} & \range{-0.511}{+0.511} & \range{-0.126}{+0.122} \\ \noalign{\smallskip}
& 18--22 & $ -0.329 $ & \range{-0.188}{+0.181} & \range{-0.358}{+0.361} & \range{-0.519}{+0.550} & \range{-0.144}{+0.145} \\
\end{tabular}
\end{table}

%% file: BPH-13-003-authorlist.tex
\textbf{Yerevan Physics Institute,  Yerevan,  Armenia}\\*[0pt]
S.~Chatrchyan, V.~Khachatryan, A.M.~Sirunyan, A.~Tumasyan
\vskip\cmsinstskip
\textbf{Institut f\"{u}r Hochenergiephysik der OeAW,  Wien,  Austria}\\*[0pt]
W.~Adam, T.~Bergauer, M.~Dragicevic, J.~Er\"{o}, C.~Fabjan\cmsAuthorMark{1}, M.~Friedl, R.~Fr\"{u}hwirth\cmsAuthorMark{1}, V.M.~Ghete, N.~H\"{o}rmann, J.~Hrubec, M.~Jeitler\cmsAuthorMark{1}, W.~Kiesenhofer, V.~Kn\"{u}nz, M.~Krammer\cmsAuthorMark{1}, I.~Kr\"{a}tschmer, D.~Liko, I.~Mikulec, D.~Rabady\cmsAuthorMark{2}, B.~Rahbaran, C.~Rohringer, H.~Rohringer, R.~Sch\"{o}fbeck, J.~Strauss, A.~Taurok, W.~Treberer-Treberspurg, W.~Waltenberger, C.-E.~Wulz\cmsAuthorMark{1}
% ...
\vskip\cmsinstskip
% ...
1:~~Also at Vienna University of Technology, Vienna, Austria\\
2:~~Also at CERN, European Organization for Nuclear Research, Geneva, Switzerland\\
% ...


%% file: BPH-13-003_temp.bbl
\providecommand{\href}[2]{#2}\begingroup\raggedright\begin{thebibliography}{10}%
\makeatletter
\providecommand{\hrefCMSnoop }[0]{\@secondoftwo}%
\makeatother
\providecommand{\doi}{\texttt{doi:}\begingroup \urlstyle{tt}\Url}

\bibitem{bib:QWG}
\hrefCMSnoop {} {N.~Brambilla {et~al.}, ``{Heavy quarkonium: progress, puzzles,
  and opportunities}'',} \textit{ Eur. Phys. J. C} \textbf{ 71} (2011) 1534,
  \href{http://dx.doi.org/10.1140/epjc/s10052-010-1534-9}{\doi{10.1140/epjc/s10052-010-1534-9}},
\href{http://www.arXiv.org/abs/1010.5827}{\texttt{ arXiv:1010.5827}}.
%%CITATION = ARXIV:1010.5827;%%.

\bibitem{bib:NRQCD}
\hrefCMSnoop {} {G.~T. Bodwin, E.~Braaten, and G.~P. Lepage, ``{Rigorous QCD
  analysis of inclusive annihilation and production of heavy quarkonium}'',}
  \textit{ Phys. Rev. D} \textbf{ 51} (1995) 1125,
  \href{http://dx.doi.org/10.1103/PhysRevD.55.5853}{\doi{10.1103/PhysRevD.55.5853}},
\href{http://www.arXiv.org/abs/hep-ph/9407339}{\texttt{ arXiv:hep-ph/9407339}}.
%%CITATION = HEP-PH/9407339;%%.

\bibitem{bib:lansberg-HP08}
\hrefCMSnoop {} {J.-P. Lansberg, ``{On the mechanisms of heavy-quarkonium
  hadroproduction}'',} \textit{ Eur. Phys. J. C} \textbf{ 61} (2009) 693,
  \href{http://dx.doi.org/10.1140/epjc/s10052-008-0826-9}{\doi{10.1140/epjc/s10052-008-0826-9}},
\href{http://www.arXiv.org/abs/0811.4005}{\texttt{ arXiv:0811.4005}}.
%%CITATION = ARXIV:0811.4005;%%.

\bibitem{bib:BK}
\hrefCMSnoop {} {M.~Beneke and M.~Kramer, ``{Direct \JPsi and $\psi^\prime$
  polarization and cross-sections at the Tevatron}'',} \textit{ Phys. Rev. D}
  \textbf{ 55} (1997) 5269,
  \href{http://dx.doi.org/10.1103/PhysRevD.55.5269}{\doi{10.1103/PhysRevD.55.5269}},
\href{http://www.arXiv.org/abs/hep-ph/9611218}{\texttt{ arXiv:hep-ph/9611218}}.
%%CITATION = HEP-PH/9611218;%%.

\bibitem{bib:Lei}
\hrefCMSnoop {} {A.~K. Leibovich, ``$\psi^\prime$ polarization due to
  color-octet quarkonia production'',} \textit{ Phys. Rev. D} \textbf{ 56}
  (1997) 4412,
  \href{http://dx.doi.org/10.1103/PhysRevD.56.4412}{\doi{10.1103/PhysRevD.56.4412}},
\href{http://www.arXiv.org/abs/hep-ph/9610381}{\texttt{ arXiv:hep-ph/9610381}}.
%%CITATION = HEP-PH/9610381;%%.

\bibitem{bib:BKL}
\hrefCMSnoop {} {E.~Braaten, B.~A. Kniehl, and J.~Lee, ``{Polarization of
  prompt \JPsi\ at the Fermilab Tevatron}'',} \textit{ Phys. Rev. D} \textbf{
  62} (2000) 094005,
  \href{http://dx.doi.org/10.1103/PhysRevD.62.094005}{\doi{10.1103/PhysRevD.62.094005}},
\href{http://www.arXiv.org/abs/hep-ph/9911436}{\texttt{ arXiv:hep-ph/9911436}}.
%%CITATION = HEP-PH/9911436;%%.

\bibitem{bib:CDFpolRun2}
\hrefCMSnoop {} {{ CDF} Collaboration, ``{Polarization of \JPsi\ and $\psi$(2S)
  mesons produced in $p \bar{p}$ collisions at $\sqrt{s}=1.96$\,TeV}'',}
  \textit{ Phys. Rev. Lett.} \textbf{ 99} (2007) 132001,
  \href{http://dx.doi.org/10.1103/PhysRevLett.99.132001}{\doi{10.1103/PhysRevLett.99.132001}},
\href{http://www.arXiv.org/abs/0704.0638}{\texttt{ arXiv:0704.0638}}.
%%CITATION = ARXIV:0704.0638;%%.

\bibitem{bib:Faccioli-EPJC}
\hrefCMSnoop {} {P.~Faccioli, C.~Louren\c{c}o, J.~Seixas, and H.~K. W{\"o}hri,
  ``Towards the experimental clarification of quarkonium polarization'',}
  \textit{ Eur. Phys. J. C} \textbf{ 69} (2010) 657,
  \href{http://dx.doi.org/10.1140/epjc/s10052-010-1420-5}{\doi{10.1140/epjc/s10052-010-1420-5}},
\href{http://www.arXiv.org/abs/1006.2738}{\texttt{ arXiv:1006.2738}}.
%%CITATION = ARXIV:1006.2738;%%.

\bibitem{bib:Faccioli-PRL-FT2Coll}
\hrefCMSnoop {} {P.~Faccioli, C.~Louren\c{c}o, J.~Seixas, and H.~K. W{\"o}hri,
  ``{\JPsi Polarization from Fixed-Target to Collider Energies}'',} \textit{
  Phys. Rev. Lett.} \textbf{ 102} (2009) 151802,
  \href{http://dx.doi.org/10.1103/PhysRevLett.102.151802}{\doi{10.1103/PhysRevLett.102.151802}},
\href{http://www.arXiv.org/abs/0902.4462}{\texttt{ arXiv:0902.4462}}.
%%CITATION = ARXIV:0902.4462;%%.

\bibitem{bib:Faccioli-PRL-FrameInv}
\hrefCMSnoop {} {P.~Faccioli, C.~Louren\c{c}o, and J.~Seixas,
  ``Rotation-invariant relations in vector meson decays into fermion pairs'',}
  \textit{ Phys. Rev. Lett.} \textbf{ 105} (2010) 061601,
  \href{http://dx.doi.org/10.1103/PhysRevLett.105.061601}{\doi{10.1103/PhysRevLett.105.061601}},
\href{http://www.arXiv.org/abs/1005.2601}{\texttt{ arXiv:1005.2601}}.
%%CITATION = ARXIV:1005.2601;%%.

\bibitem{bib:Faccioli-PRD-FrameInv}
\hrefCMSnoop {} {P.~Faccioli, C.~Louren\c{c}o, and J.~Seixas, ``New approach to
  quarkonium polarization studies'',} \textit{ Phys. Rev. D} \textbf{ 81}
  (2010) 111502(R),
  \href{http://dx.doi.org/10.1103/PhysRevD.81.111502}{\doi{10.1103/PhysRevD.81.111502}},
\href{http://www.arXiv.org/abs/1005.2855}{\texttt{ arXiv:1005.2855}}.
%%CITATION = ARXIV:1005.2855;%%.

\bibitem{bib:Faccioli-shapes}
\hrefCMSnoop {} {P.~Faccioli, C.~Louren\c{c}o, J.~Seixas, and H.~K. W{\"o}hri,
  ``Model-independent constraints on the shape parameters of dilepton angular
  distributions'',} \textit{ Phys. Rev. D} \textbf{ 83} (2011) 056008,
  \href{http://dx.doi.org/10.1103/PhysRevD.83.056008}{\doi{10.1103/PhysRevD.83.056008}},
\href{http://www.arXiv.org/abs/1102.3946}{\texttt{ arXiv:1102.3946}}.
%%CITATION = ARXIV:1102.3946;%%.

\bibitem{PhysRevLett.108.151802}
\hrefCMSnoop {} {{ CDF} Collaboration, ``{Measurements of Angular Distributions
  of Muons from $\Upsilon$ Meson Decays in $p\bar{p}$ Collisions at
  $\sqrt{s}=1.96$\,TeV}'',} \textit{ Phys. Rev. Lett.} \textbf{ 108} (2012)
  151802,
  \href{http://dx.doi.org/10.1103/PhysRevLett.108.151802}{\doi{10.1103/PhysRevLett.108.151802}},
\href{http://www.arXiv.org/abs/1112.1591}{\texttt{ arXiv:1112.1591}}.
%%CITATION = ARXIV:1112.1591;%%.

\bibitem{Baranov:2011ib}
\hrefCMSnoop {} {S.~P. Baranov, A.~V. Lipatov, and N.~P. Zotov, ``{Prompt \JPsi
  production at LHC: new evidence for the $k_T$-factorization}'',} \textit{
  Phys. Rev. D} \textbf{ 85} (2012) 014034,
  \href{http://dx.doi.org/10.1103/PhysRevD.85.014034}{\doi{10.1103/PhysRevD.85.014034}},
\href{http://www.arXiv.org/abs/1108.2856}{\texttt{ arXiv:1108.2856}}.
%%CITATION = ARXIV:1108.2856;%%.

\bibitem{bib:UpsPol-CMS}
\hrefCMSnoop {} {{ CMS} Collaboration, ``{Measurement of the $\Upsilon$(1S),
  $\Upsilon$(2S), and $\Upsilon$(3S) polarizations in pp collisions at
  $\sqrt{s} = 7$\,TeV}'',} \textit{ Phys. Rev. Lett.} \textbf{ 110} (2013)
  081802,
  \href{http://dx.doi.org/10.1103/PhysRevLett.110.081802}{\doi{10.1103/PhysRevLett.110.081802}},
  \href{http://www.arXiv.org/abs/1209.2922}{\texttt{ arXiv:1209.2922}}.

\bibitem{bib:psiPol-ALICE}
\hrefCMSnoop {} {{ ALICE} Collaboration, ``{\JPsi polarization in $pp$
  collisions at $\sqrt{s}=7$ TeV}'',} \textit{ Phys. Rev. Lett.} \textbf{ 108}
  (2012) 082001,
  \href{http://dx.doi.org/10.1103/PhysRevLett.108.082001}{\doi{10.1103/PhysRevLett.108.082001}},
\href{http://www.arXiv.org/abs/1111.1630}{\texttt{ arXiv:1111.1630}}.
%%CITATION = ARXIV:1111.1630;%%.

\bibitem{bib:psiPol-LHCb}
\hrefCMSnoop {} {{ LHCb} Collaboration, ``{Measurement of $J/\psi$ polarization
  in $pp$ collisions at $\sqrt{s}=7$ TeV}'',} (2013).
\href{http://www.arXiv.org/abs/1307.6379}{\texttt{ arXiv:1307.6379}}.
%%CITATION = ARXIV:1307.6379;%%.

\bibitem{bib:Faccioli-feeddown}
\hrefCMSnoop {} {P.~Faccioli, C.~Louren\c{c}o, J.~Seixas, and H.~K. W{\"o}hri,
  ``{Study of $\psi^\prime$ and $\chi_{\rm c}$ decays as feed-down sources of
  \JPsi\ hadro-production}'',} \textit{ JHEP} \textbf{ 10} (2008) 004,
  \href{http://dx.doi.org/10.1088/1126-6708/2008/10/004}{\doi{10.1088/1126-6708/2008/10/004}},
\href{http://www.arXiv.org/abs/0809.2153}{\texttt{ arXiv:0809.2153}}.
%%CITATION = ARXIV:0809.2153;%%.

\bibitem{bib:CS}
\hrefCMSnoop {} {J.~C. Collins and D.~E. Soper, ``{Angular Distribution of
  Dileptons in High-Energy Hadron Collisions}'',} \textit{ Phys. Rev. D}
  \textbf{ 16} (1977) 2219,
\href{http://dx.doi.org/10.1103/PhysRevD.16.2219}{\doi{10.1103/PhysRevD.16.2219}}.
%%CITATION = PHRVA,D16,2219;%%.

\bibitem{Braaten:2008mz}
\hrefCMSnoop {} {E.~Braaten, D.~Kang, J.~Lee, and C.~Yu, ``{Optimal spin
  quantization axes for the polarization of dileptons with large transverse
  momentum}'',} \textit{ Phys. Rev. D} \textbf{ 79} (2009) 014025,
  \href{http://dx.doi.org/10.1103/PhysRevD.79.014025}{\doi{10.1103/PhysRevD.79.014025}},
\href{http://www.arXiv.org/abs/0810.4506}{\texttt{ arXiv:0810.4506}}.
%%CITATION = ARXIV:0810.4506;%%.

\bibitem{Chatrchyan:2008zzk}
\hrefCMSnoop {} {{ CMS} Collaboration, ``The {CMS} experiment at the {CERN}
  {LHC}'',} \textit{ JINST} \textbf{ 03} (2008) S08004,
  \href{http://dx.doi.org/10.1088/1748-0221/3/08/S08004}{\doi{10.1088/1748-0221/3/08/S08004}}.

\bibitem{Khachatryan:2010xn}
\hrefCMSnoop {} {{ CMS} Collaboration, ``{Measurements of Inclusive W and Z
  Cross Sections in pp Collisions at $\sqrt{s}=7$\,TeV}'',} \textit{ JHEP}
  \textbf{ 01} (2011) 080,
  \href{http://dx.doi.org/10.1007/JHEP01(2011)080}{\doi{10.1007/JHEP01(2011)080}},
\href{http://www.arXiv.org/abs/1012.2466}{\texttt{ arXiv:1012.2466}}.
%%CITATION = ARXIV:1012.2466;%%.

\bibitem{CrystalBall}
\href {http://www.slac.stanford.edu/pubs/slacreports/slac-r-236.html} {M.~J.
  Oreglia, ``A study of the reactions $\psi^\prime \to \gamma \gamma \psi$''}.
\newblock PhD thesis, Stanford University, 1980.
\newblock {SLAC} Report {SLAC-R-236}.

\bibitem{bib:PDG2012}
\hrefCMSnoop {} {{Particle Data Group}, J.~Beringer {et~al.}, ``{Review of
  Particle Physics}'',} \textit{ Phys. Rev. D} \textbf{ 86} (2012) 010001,
  \href{http://dx.doi.org/10.1103/PhysRevD.86.010001}{\doi{10.1103/PhysRevD.86.010001}}.

\bibitem{bib:BPH-10-002}
\hrefCMSnoop {} {{ CMS} Collaboration, ``{Prompt and non-prompt \JPsi
  production in $pp$ collisions at $\sqrt{s}=7$\,TeV}'',} \textit{ Eur. Phys.
  J. C} \textbf{ 71} (2011) 1575,
  \href{http://dx.doi.org/10.1140/epjc/s10052-011-1575-8}{\doi{10.1140/epjc/s10052-011-1575-8}},
\href{http://www.arXiv.org/abs/1011.4193}{\texttt{ arXiv:1011.4193}}.
%%CITATION = ARXIV:1011.4193;%%.

\bibitem{Gong:2012ug}
\hrefCMSnoop {} {B.~Gong, L.-P. Wan, J.-X. Wang, and H.-F. Zhang,
  ``{Polarization for Prompt \JPsi, $\psi(2S)$ production at the Tevatron and
  LHC}'',} \textit{ Phys. Rev. Lett.} \textbf{ 110} (2013) 042002,
  \href{http://dx.doi.org/10.1103/PhysRevLett.110.042002}{\doi{10.1103/PhysRevLett.110.042002}},
\href{http://www.arXiv.org/abs/1205.6682}{\texttt{ arXiv:1205.6682}}.
%%CITATION = ARXIV:1205.6682;%%.

\bibitem{Butenschoen:2012px}
\hrefCMSnoop {} {M.~Butenschoen and B.~A. Kniehl, ``{\JPsi polarization at
  Tevatron and LHC: Nonrelativistic-QCD factorization at the crossroads}'',}
  \textit{ Phys. Rev. Lett.} \textbf{ 108} (2012) 172002,
  \href{http://dx.doi.org/10.1103/PhysRevLett.108.172002}{\doi{10.1103/PhysRevLett.108.172002}},
\href{http://www.arXiv.org/abs/1201.1872}{\texttt{ arXiv:1201.1872}}.
%%CITATION = ARXIV:1201.1872;%%.

\bibitem{Chao:2012iv}
K.-T. Chao\hrefCMSnoop {} { {et~al.}, ``{\JPsi Polarization at Hadron Colliders
  in Nonrelativistic QCD}'',} \textit{ Phys. Rev. Lett.} \textbf{ 108} (2012)
  242004,
  \href{http://dx.doi.org/10.1103/PhysRevLett.108.242004}{\doi{10.1103/PhysRevLett.108.242004}},
\href{http://www.arXiv.org/abs/1201.2675}{\texttt{ arXiv:1201.2675}}.
%%CITATION = ARXIV:1201.2675;%%.

\end{thebibliography}\endgroup
